\documentclass[prd,aps,preprintnumbers,nofootinbib,twocolumn]{revtex4-1}
\usepackage{quiver}
\usepackage{amscd}
\usepackage{amsmath, bbm} 
\usepackage{amsfonts}
\usepackage{amssymb}
\usepackage{slashed}
\usepackage{multirow}
\usepackage{array}
\usepackage{cancel}
\usepackage{xcolor}
\usepackage{hyperref}
\usepackage{axodraw2}
\usepackage{tikz-cd}
\usetikzlibrary{arrows,decorations.markings}
\usepackage{enumitem}
\usepackage[normalem]{ulem}
\usepackage[export]{adjustbox}
\usepackage{placeins} % provides \FloatBarrier

\usepackage{listings}% http://ctan.org/pkg/listings
\usepackage{tabularx}

\newcommand{\R}{\mathbb{R}}

\newcommand{\su}{\mathfrak{su}}
\newcommand{\so}{\mathfrak{so}}
\newcommand{\spalg}{\mathfrak{sp}}
\renewcommand{\r}{\mathfrak{r}}
\renewcommand{\a}{\mathfrak{a}}
\newcommand{\s}{\mathfrak{s}}
\renewcommand{\u}{\mathfrak{u}(1)}
\newcommand{\g}{\mathfrak{g}}
\newcommand{\h}{\mathfrak{h}}

\newcommand{\SMnu}{$\mathrm{SM}_\nu$}

\makeatletter
\renewcommand{\p@subsection}{}

\makeatother

\begin{document} 
\title{Superfloccinaucinihilipilification:\\
Semisimple unifications of any gauge theory}
\author{Andrew Gomes} \email{awg76@cornell.edu} \affiliation{Laboratory for Elementary Particle Physics, Cornell University, Ithaca, NY 14853, USA}
\author{Maximillian Ruhdorfer} \email{m.ruhdorfer@cornell.edu} \affiliation{Laboratory for Elementary Particle Physics, Cornell University, Ithaca, NY 14853, USA}
\author{Joseph Tooby-Smith} \email{j.tooby-smith@cornell.edu} \affiliation{Laboratory for Elementary Particle Physics, Cornell University, Ithaca, NY 14853, USA}
\begin{abstract}
We present a Mathematica package that takes any reductive gauge algebra and fully-reducible fermion representation, and outputs all semisimple gauge extensions under the condition that they have no additional fermions, and are free of local anomalies. These include all simple completions, also known as grand unified theories (GUT). We additionally provide a list of all semisimple completions for 5835 fermionic extensions of the one-generation Standard Model.
\end{abstract}
\maketitle
\tableofcontents

%%%%%%%%%%%%%%%%%
\section{Introduction}
%%%%%%%%%%%%%%%%%

Unification, the idea that the Standard Model (SM) gauge algebra $\su (3)\oplus \su (2) \oplus \u$  and particle representations embed into a semisimple algebra in the UV, is an appealing scenario for physics beyond the Standard Model (BSM). Theories based on semisimple algebras have phenomenological benefits over their reductive counterparts (those algebras containing $\u$ factors). These include the simplicity of local anomaly-cancellation, a possible origin for flavor symmetries, and the freedom from Landau poles.

Arguably the most famous of such extensions are Grand Unified Theories (GUTs), a particularly elegant subclass where the unifying algebra is simple. These include the well-known  $\mathfrak{su}(5)$ and $\mathfrak{so}(10)$ GUTs~\cite{Georgi:1974sy,Fritzsch:1974nn,Georgi:1974my}, which unify the fermion content of the SM ---including right-handed neutrinos (RHN) in the case of $\so (10)$--- into representations of $\mathfrak{su}(5)$ and $\mathfrak{so}(10)$, respectively. Another theoretically well-motivated subclass are semisimple extensions which mix flavor with gauge symmetries in a non-trivial way. Examples are those based on the Pati-Salam $\su(4)\oplus \su(2)\oplus\su(2)$ model~\cite{Pati:1974yy} such as ``color-flavor" unification (e.g. $\mathfrak{su}(12)\oplus \mathfrak{su}(2)\oplus \su(2)$) and ``electroweak-flavor" unification (e.g. $\su(4)\oplus \mathfrak{sp}(6)\oplus \mathfrak{sp}(6)$)~\cite{Davighi:2022fer}. 

In the past, searching for semisimple extensions of a model was a tedious process that involved manual calculations and guesswork. The approach developed in~\cite{Allanach:2021bfe}, which we will refer to as ``Flocci'' (short for floccinaucinihilipilification), produced a complete list of the 340 possible anomaly-free semisimple extensions of the three generation Standard Model with right-handed neutrinos (\SMnu). However, Flocci was specifically tailored to the SM gauge algebra and fermion content and was not able to deal with different gauge algebras or a modified fermion content.

In this paper, we rectify this shortcoming and introduce ``superfloccinaucinihilipilification" (or ``SuperFlocci"), a Mathematica package that implements a more comprehensive and fundamentally different approach that can handle any reductive Lie algebra and any fully-reducible anomaly-free fermion representation as an input. SuperFlocci outputs all semisimple anomaly-free extensions with a fixed number of fermions, including the maximal and minimal extensions and the branching patterns between them.

Like Flocci, the underlying workings of SuperFlocci rely heavily on the theory of Lie algebras, in particular the references~\cite{de2000lie,Lorente:1972xw,gruber1975semisimple,dyn52:MaximalSubgroups,dynkin1952semisimple,Feger:2019tvk}. However, SuperFlocci uses a much more streamlined approach, building a tree of semisimple extensions starting from the input. Each edge of the tree corresponds to a maximal embedding of semisimple Lie algebras. Flocci, on the other hand, took a much more computationally expensive brute force approach that was difficult to generalize. 

SuperFlocci has many applications, including:
\begin{itemize}
\item Determining the existence of a semisimple completion for a given theory.
\item Finding possible flavor symmetries of a given fermion content and gauge algebra, whether horizontal or mixed with the input gauge algebra.
\item Helping identify possible breaking patterns from a given semisimple completion down to the input algebra, through other semisimple completions.
\item Determining the semisimple completions of $\u$-extensions of a theory.
\item Scanning over theories of different fermion content, and searching for interesting semisimple extensions of each one.
\end{itemize}

It is worth noting that SuperFlocci only considers the fermionic sector, i.e. it cannot provide any insight into questions which require information about the scalar degrees of freedom. This includes the running of couplings, how a symmetry-breaking pattern is achieved through scalar vacuum expectation values, and the question of whether and at what scale the gauge couplings unify. However, the output of SuperFlocci is an exhaustive list of all possible unified models and can serve as a starting point for model building of a realistic unified theory.

The paper is organized as follows. We start with an explanation of how to download, run, and interpret the output of SuperFlocci in Section~\ref{sec:running}. Section~\ref{sec:inout} describes formally the program's inputs and outputs. In Section~\ref{sec:checkscan} we perform a number of checks, showing the agreement of SuperFlocci with previous results, and perform a scan over BSM theories, looking for interesting semisimple extensions. Section~\ref{sec:example} provides an intuitive description of SuperFlocci's underlying algorithm with the help of a worked-out example. The Appendices contain our conventions for Dynkin diagrams and summaries of the results of the example runs of Section~\ref{sec:checkscan}.

%%%%%%%%%%%%%%%%%
\section{Running the code}\label{sec:running}
%%%%%%%%%%%%%%%%%
%
We have designed SuperFlocci so that it is easy to use, with minimal prerequisites. It has been tested on Mathematica v12.

The most up-to-date version of our code can be downloaded from GitHub at
	\begin{multline*}
		\href{https://github.com/jstoobysmith/Superfloccinaucinihilipilification}{\mathtt{https://github.com/jstoobysmith}}\\
		\href{https://github.com/jstoobysmith/Superfloccinaucinihilipilification}{\mathtt{/Superfloccinaucinihilipilification}}
	\end{multline*}
To install, download the file \verb|SuperFlocci.m|, and import it into a Mathematica notebook using:
\begin{verbatim}
Import["SuperFlocci.m"];
\end{verbatim}

To find all semisimple extensions (within our criterion) we call the \verb|SuperFlocci| function as
\begin{verbatim}
	SuperFlocci[alg,rep]
\end{verbatim}
which takes the required arguments 
 \begin{itemize}
	\item[\texttt{alg:}] a reductive gauge algebra in the form of a list of simple or $\u$ factors labeled according to the Dynkin classification, e.g. \verb|{A1,A2,U1}| for $\su(2)\oplus\su(3)\oplus\u$.
    \item[\texttt{rep:}] a fermion representation in the form of a list of highest weights in the same order as the algebra, e.g. the highest weight of the $(\mathbf{2},\mathbf{3})_1$ representation is written as \verb|{1,1,0,1}|. Charges associated with $\u$ factors should be scaled to integers.
\end{itemize}
A list of optional arguments is given in Table~\ref{tab:flocciargs}.

SuperFlocci uses Dynkin notation to denote simple Lie algebras. A conversion of the Dynkin notation for the classical Lie algebras to the more commonly used notation in physics, together with their syntax in SuperFlocci, is given in Table~\ref{tab:LieAlgebras}.  Our convention (though not essential in the use of SuperFlocci) will be to write algebras with simple factors of smaller rank first, and with $\u$ factors at the end. For example the SM gauge algebra is $\su(2) \oplus \su(3) \oplus \u$.
\begin{table}[h]
\renewcommand{\arraystretch}{1.5} 
\setlength{\tabcolsep}{12pt}
	\begin{tabular}{ccc}
	\hline
		{\bf Dynkin}	&	{\bf Conventional}	&	{\bf \verb|SuperFlocci|}\\
		\hline
		$A_n$	&	$\mathfrak{su}(n+1)$		&	\verb|An|\\
		$B_n$	&	$\mathfrak{so}(2n+1)$	&	\verb|Bn|\\
		$C_n$	&	$\mathfrak{sp}(2n)$		&	\verb|Cn|\\
		$D_n$	&	$\mathfrak{so}(2n)$		&	\verb|Dn|\\
		$E_6, E_7, E_8$	& $E_6, E_7, E_8$ & \verb|E6,E7,E8|\\
		$F_4$			& $F_4$			& \verb|F4|\\
		$G_2$			& $G_2$			& \verb|G2|\\
		-&$\u$ & \verb|U1|\\
		\hline
	\end{tabular}
	\caption{Notation for the classical Lie algebras, and $\u$, and their syntax in SuperFlocci.}
	\label{tab:LieAlgebras}
\end{table}

The weights follow the ordering convention of simple roots as indicated in the Dynkin diagrams of Appendix~\ref{app:DynkinDiagrams}. For an encyclopedic reference on the representations of Lie algebras and their expression in highest weight form, see e.g.~\cite{Feger:2019tvk}.  Here are some of the most common ones: 
\begin{equation}
	\begin{aligned}[t]
		\su(2):\quad 	& \mathbf{1} = (0) \\ 
  						& \mathbf{2} = (1) \\  
  						& \mathbf{3} = (2)
	\end{aligned}\qquad\qquad
	\begin{aligned}[t]
		\su(3):\quad 	& \mathbf{1} = (0,0) \\ 
  						& \mathbf{3} = (1,0) \\  
  						& \mathbf{\overline{3}} = (0,1) \\  
  						& \mathbf{6} = (2,0) \\
  						& \overline{\mathbf{6} }= (0,2) \\
  						& \mathbf{8} = (1,1) 
	\end{aligned}
\end{equation}

As an explicit example, consider the one-generation Standard Model with a right-handed neutrino (which we shall refer to as $\text{SM}_\nu$ throughout). This has the fermion representation $Q\oplus U\oplus D\oplus L\oplus E\oplus N$, where
\begin{align}
Q &= (\mathbf{2},\mathbf{3})_1 \nonumber \\
U &= (\mathbf{1},\mathbf{\overline{3}})_{-4} \nonumber \\
D &= (\mathbf{1},\mathbf{\overline{3}})_2 \label{eq:qudlen}\\
L &= (\mathbf{2},\mathbf{1})_{-3} \nonumber \\
E &= (\mathbf{1},\mathbf{1})_6 \nonumber \\
N &= (\mathbf{1},\mathbf{1})_0 .\nonumber 
\end{align}
To run SuperFlocci on this example use:
\begin{verbatim}
out = SuperFlocci[{A1,A2,U1}, {{1,1,0,1},
        {0,0,1,-4},{0,0,1,2},{1,0,0,-3},
        {0,0,0,6},{0,0,0,0}}];
\end{verbatim}
The highest weights here correspond respectively to the irreducible representations (irreps) in (\ref{eq:qudlen}). On a standard laptop, this example takes only a few seconds to run. In comparison, the three generation $\text{SM}_\nu$ takes of order an hour, similar to the time taken by Flocci (although SuperFlocci provides more information). 

All the information we might need is contained in the \verb|out| variable. This includes the input algebra, the input representation,  semisimple extensions (including projections $\Lambda \kappa$ to the input algebra),  projection matrices $\Lambda \rho$ between extensions,  and timing information. However, for ease of reading we can use the \verb|FlocciOutput| function:
\begin{verbatim}
FlocciOutput[out];
\end{verbatim}

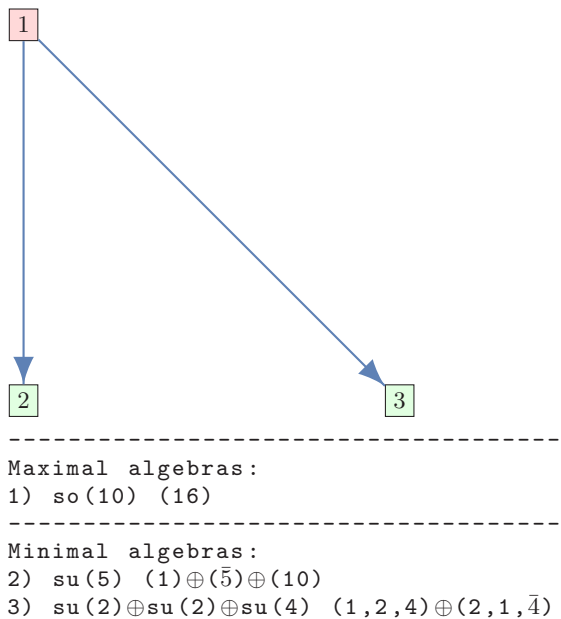
\begin{figure}[h]
\definecolor{lightred}{rgb}{1, 0.85, 0.85}
\definecolor{lightgreen}{rgb}{0.88, 1, 0.88}
\definecolor{lightblue}{rgb}{0.368417,0.506779,0.709798}
\begin{adjustbox}{padding*=-14ex 0ex 0ex 0ex}
\begin{tikzpicture}[node distance={50mm}, thick, main/.style = {}] 
\node[main,draw, thin,fill=lightred] (1) {1}; 
\node[main,draw, thin, fill=lightgreen] (2) [below of=1] {2}; 
\node[main,draw, thin, fill=lightgreen] (3) [right of=2] {3}; 
\draw [lightblue,decoration={markings,mark=at position 1 with
    {\arrow[scale=2,>=latex]{>}}},postaction={decorate}]  (1) -- (2); 
\draw [lightblue,decoration={markings,mark=at position 1 with
    {\arrow[scale=2,>=latex]{>}}},postaction={decorate}] (1) -- (3); 
\end{tikzpicture} 
\end{adjustbox}
\begin{adjustbox}{center}
\begin{lstlisting}
-------------------------------------
Maximal algebras:
1) so(10) (16)
-------------------------------------
Minimal algebras: 
2) su(5) (1)$\oplus$($\bar 5$)$\oplus$(10)
3) su(2)$\oplus$su(2)$\oplus$su(4) (1,2,4)$\oplus$(2,1,$\bar 4$)
\end{lstlisting}
\end{adjustbox}
 \caption{The output of SuperFlocci, with the one-generation $\text{SM}_\nu$ as input.}
\label{fig:screen}
\end{figure}

This will display a graph (size permitting) of all semisimple extensions with edges representing embeddings, and lists of the maximal and minimal extensions (see Section \ref{sec:output} for an explanation of these features). Figure \ref{fig:screen} shows the output for this example.

\begin{table*}
\renewcommand{\arraystretch}{1.5} 
\fontsize{8}{9.2}\selectfont

\centering
  
  \begin{tabularx}{18cm}{|l|l|X|}
    \hline
    \textbf{Argument} & \textbf{Type (Default)} & \textbf{Description} \\
    \hline
    \verb|Checkpoint| & String (null) & The filename where checkpointed data should be saved. Can be used for long computations in case of a crash. \\
    \hline
    \verb|CheckpointUpFreq| & Integer (1000) & How frequently (measured in nodes) checkpoints should be carried out in the upward growth phase. \\
    \hline   
    \verb|StartFromCheckpoint| & Boolean (True) & If a checkpoint filename is given, and the file exists, start from saved checkpoint. \\
    \hline
    \verb|ClearDataFreq| & Integer (1000) & How often (measured in nodes) Mathematica cache should be cleared. \\
    \hline
    \verb|SimpleIdealConstraint| & Integer ($\infty$) & An upper limit on the number of simple ideals of nodes generated. \\
    \hline
    \verb|DetailedProgressData| & Boolean (False) & Display detailed progress data (e.g. memory usage) while the code is running. \\
     \hline
    \verb|ExtendedKappaCheck| & Boolean (True) & Determines whether or not pruning described in Section~\ref{sec:pruning} is carried out. \\
    \hline
  \end{tabularx}
  \caption{The optional arguments of \texttt{SuperFlocci}.}
  \label{tab:flocciargs}
\end{table*}

\begin{table*}
\renewcommand{\arraystretch}{1.5} 
\fontsize{8}{9.2}\selectfont
  \centering
  \begin{tabularx}{18cm}{|l|l|X|}
    \hline
    \textbf{Argument} & \textbf{Type (Default)} & \textbf{Description} \\
    \hline
    \verb|display| & Boolean (True) & Whether to display the output. \\
    \hline
    \verb|save| & Boolean (False) & Whether to save the output to a file. \\
    \hline   
    \verb|filename| & String (null) & Where to save the output (without an extension - it will be saved as a .m file). \\
    \hline
    \verb|latex| & String list (null) & 
    Create a LaTeX file (with a default name) with customizable information given by the following options:  
    \begin{enumerate}[wide=5pt, leftmargin=*] \item[\texttt{"maxtables"}:] table with all maximal algebras.  \item[\texttt{"mintables"}:] table with all minimal algebras.   \item[\texttt{"alltables"}:] tables with all algebras.  \item[\texttt{"embeddingsdetail"}:] detailed overview of all embeddings.  \item[\texttt{"tensorproduct"}:] (requires \verb|"embeddingsdetail"|): gives all bilinear tensorproducts of representations.  \item[\texttt{"onlymaxminembeddings"}:] (requires \verb|"embeddingsdetail"|): restricts \verb|"embeddingsdetail"| to maximal and minimal embeddings.  \item[\texttt{"projectiontree"}:] (requires \verb|"embeddingsdetail"|): outputs projection matrices between algebras on same branch of tree (is neglected for \verb|"onlymaxminembeddings"|).\end{enumerate} \\
    \hline
    \verb|labelrep| & Rule list (null) & Replacement rules for highest weights to increase readability of LaTeX output. For example, use the rule list \verb|{{1,1,0,1}->"Q"}| to replace the left-handed quark representation with the letter Q. \\
    \hline
  \end{tabularx}
  
   \caption{The optional arguments of \texttt{FlocciOutput}.}
  \label{tab:outargs}
\end{table*}

The optional arguments of \verb|FlocciOutput| are given in Table \ref{tab:outargs}. Suppose that, in addition to displaying the output, we want to save it to the file \verb|output.m| and write LaTeX code to display all generated algebras and embeddings. We would accomplish this with:
\begin{verbatim}
FlocciOutput[out, save -> True,
  filename -> "output", latex -> 
  {"alltables","embeddingsdetail"}];
\end{verbatim}

Examples of the tables generated by the LaTeX output can be found in Appendix \ref{app:tables}.  {The embedding data comes in the form of the projection matrix $\Lambda \kappa$ to the input algebra and the projection matrices $\Lambda \rho$ to the maximal subalgebras (see Figure~\ref{fig:Maps})}. Applying these matrices to the weight system (generated by the highest weights) of a particular semisimple extension, one obtains the weight system of the input representation and the weight systems of all maximal subalgebras, respectively.

While SuperFlocci can in principle handle any input data conforming to the requirements of Section \ref{sec:inout}, certain features of the input require more computation time or memory. Large input representations, vector-like sectors in the input, and singlets are all factors that lead to more anomaly free combinations of irreps, and thus more extensions to be checked. Additionally, as the number of distinct charges assigned to a given weight (coming from different irreps that share a semisimple weight) increases, the program slows with a corresponding combinatoric factor.  {For example, we were able to run 5 generations of the $\su(5)$ GUT fermion content $(\mathbf{\bar{5}}\oplus \mathbf{10})^{\oplus 5}$ on a standard laptop, but not the 4 generation SM.} Finally, any computation involving the high-rank exceptional algebras will take longer due to their many automorphisms, which must be hard-coded, in contrast with the automorphisms of high-rank classical Lie algebras.

%%%%%%%%%%%%%%%%%
\section{Description of the input and output}\label{sec:inout}
%%%%%%%%%%%%%%%%%
Colloquially, our program finds all semisimple extensions of a given input Lie algebra and fermion representation. In this section, we want to make this statement more precise by carefully describing the allowed inputs to the function \verb|SuperFlocci|, and its output.

Throughout this section and Section~\ref{sec:example}, we will assume basic knowledge of the theory of Lie algebras.\footnote{Namely the notions of: semisimple and reductive Lie algebras, Cartan subalgebras, automorphisms, (highest) weights, and projections matrices. Those readers who need a refresher are directed to~\cite{Tooby-Smith:2021hjk} and references therein.} We note here that all Lie algebras throughout this paper are implicitly assumed to be complexified. 
%%%%%%%
\subsection{The input}
%%%%%%%
The input of our program consists of
\begin{itemize}
	\item A finite-dimensional reductive Lie algebra $\r$.
	\item The highest weights of a fully-reducible, $n$-dimensional representation $r:\r\to \su(n)$.\footnote{Note that as complexified Lie algebras $\mathfrak{gl}(n)$ and $\mathfrak{u}(n)$ are isomorphic. The condition of anomaly-freeness ensures $r$ factors through $\su(n)$. One can also argue for $\mathfrak{u}(n)$ using the invariance of the kinetic term.} 
\end{itemize}
The condition of ``full reducibility" here is to prevent representations of $\r$ akin to the non-diagonalizable representation of $\u$ given by 
\begin{align}
	\u\to \su(2)\,:\, \lambda \mapsto \begin{pmatrix}
		0 & \lambda\\
		0 & 0
	\end{pmatrix} .
\end{align} 

We can split the reductive Lie algebra $\r$ into the direct sum $\s\oplus \a$ of a semisimple part $\s$ and an abelian part $\a$. Let $\h_{\s}$ be a chosen Cartan subalgebra of $\s$ and $\h_{\s}^\ast$  its dual. The highest weights of $r$ as inputted into  \verb|SuperFlocci|  must be written with respect to a given basis, which we now describe. For $\h_{\s}^\ast$ we use the basis of fundamental weights with indices as indicated in the Dynkin diagrams of Appendix~\ref{app:DynkinDiagrams}. 

We also need a basis for the dual of $\a$, denoted $\a^\ast$. To do this, we place the following condition on our input 
\begin{itemize}
	\item A basis must exist of $\a^\ast$, which when combined with the basis of $\h_{\s}^\ast$, leads to integer highest weights.
\end{itemize} 
Any such basis can then be used as a valid basis of $\a$. It is for this reason that the SM hypercharges in Eq.~(\ref{eq:qudlen}) have been scaled to be integers. This condition is satisfied if we assume that $\a$ is the algebra of a compact Lie group ({e.g.} $U(1)$ rather than $\R$).
%\js{Cite in above: Clifford Algebras and Lie Theory}

%%%%%%%
\subsection{The output}\label{sec:output}
%%%%%%%
To explain the output of SuperFlocci, we borrow some basic terminology from category theory. A category consists of objects ({e.g.} vector spaces) and morphisms ({e.g.} linear maps) between the objects, such that every object has an identity morphism and morphisms can be associatively composed.

The category we need in order to understand the output has objects which are triples $(\g, \{\beta_i\}, \Lambda \kappa)$ where  
  \begin{itemize}
  \item $\g$ is a semisimple Lie algebra,
  \item $\{\beta_i\}$ are the highest weights of a $n$-dimensional representation,
  \item $\Lambda\kappa:\h_\g^\ast\to \h_\r^\ast$ is a linear transformation (represented by a matrix).
  \end{itemize}
  subject to the conditions
  \begin{enumerate}
  	\item $\Lambda\kappa$ must extend to a complete embedding of $\r$ into $\g$.
  	\item The representation indicated  $\{\beta_i\}$ must branch to $r$ under $\Lambda \kappa$ (up-to equivalence)
  	\item The representation $\{\beta_i\}$ must be free of local anomalies.
  \end{enumerate}
A morphism between two triples $(\g,\Lambda\kappa,\{\beta_i\})\to(\g^\prime,\Lambda\kappa^\prime,\{\beta_i^\prime\})$ is a linear map $\Lambda \rho: \h_{\g}\to \h_{\g^\prime}$ such that there exists a full embedding $\rho: \g^\prime \to \g$, $\Lambda\kappa=\Lambda\kappa^\prime \Lambda \rho$, and on branching $\beta$ becomes $\beta^\prime$. Those morphisms which have an inverse under composition are called \emph{isomorphisms}. In our case, the isomorphisms are those $\Lambda \rho$ for which $\g$ is isomorphic to $\g^\prime$.
  
The output of  \verb|SuperFlocci| is \emph{not} this category. Instead, it is a directed acyclic graph (also known as a polytree). For simplicity, we shall refer to it as a tree or graph. For each isomorphism class of objects in the category we choose a representative, and these representatives form the nodes of the graph. The edges of the graph are a minimal subset of morphisms such that when composed with isomorphisms, they generate the whole category. An example of a polytree generated in this way can be found in Figure~\ref{fig:final}.

The sources of this graph (i.e. those nodes with no edges into them) are the maximal extensions, and the sinks of this graph (i.e. those nodes with no edges out of them) are the minimal extensions.
  
%%%%%%%%%%%%%%%%%
\section{Checks and examples}\label{sec:checkscan}
%%%%%%%%%%%%%%%%%
%
In this section, we demonstrate that  \verb|SuperFlocci|  is able to reproduce and extend well-known results. We present the results of running SuperFlocci on a series of BSM models, we also summarize the results of a scan over possible extensions of the SM. The Mathematica output for each of the examples in the text can be found in the GitHub repository~\cite{github}.

%
%%%%%%%
\subsection{Checks}
%%%%%%%
%
A number of obvious checks can be performed on SuperFlocci by using an input where part of the output is known beforehand. This could be, for example, from a previously known unifying theory. For these checks, we stress that the output will include much more than the sought-after unified theory. It will be a complete list of all possible semisimple extensions of the input gauge algebra and particle spectrum as well as the corresponding projection matrices and all branching patterns between them.

In what follows we go through the various checks of this sort we performed on SuperFlocci.

~\\~\\
\textbf{SM:} It is well-known that one generation of SM fermions fits into the $\bar{\mathbf{5}}$ and $\mathbf{10}$ representations of $SU(5)$~\cite{Georgi:1974sy}. If instead we consider SM$_\nu$, there are additional embeddings into $\mathfrak{su}(2)\oplus\mathfrak{su}(2)\oplus\mathfrak{su}(4)$~\cite{Pati:1974yy} and $\mathfrak{so}(10)$~\cite{Fritzsch:1974nn,Georgi:1974my}. In Section~\ref{sec:example} we use SuperFlocci to demonstrate not only the existence of these extensions, but also that these are the only semisimple extensions.

For two and three generations of SM$_\nu$, Ref.~\cite{Allanach:2021bfe} found that there are $45$ and $340$ different extensions with $9$ and $24$ of them being maximal, respectively. Despite using a different approach, SuperFlocci exactly reproduces these results and additionally finds the embeddings between these extensions.

We further ran SuperFlocci on the three-generation SM (without additional RHNs). In this case there are $19$ embeddings with $5$ of them being maximal (see Table~\ref{tab:MinMaxSMwoRHN}). These are all based on $\su(5)$. \\~\\
\textbf{Non-SM gauge algebra:} SuperFlocci is not limited to the SM gauge algebra $\mathfrak{su}(2)\oplus \mathfrak{su}(3)\oplus \mathfrak{u}(1)$ but can be used with any reductive Lie algebra. As a simple check we took the $19$ semisimple extensions of the SM without RHNs as an input and added three singlets. An example would be to take as input the algebra $\mathfrak{su}(5)$ with the fermion representation $\mathbf{\bar{5}}^{\oplus 3} \oplus \mathbf{10}^{\oplus 3}\oplus \mathbf{1}^{\oplus 3}$. We checked that the result was a strict subset of the $340$ semisimple extensions of SM$_\nu$.
\\~\\
\textbf{Extending the SM fermion spectrum:} Instead of choosing a different gauge algebra we can extend the fermion spectrum. As has been known for a long time, if one extends the one-generation SM by fermions in the vector-like representations $(\mathbf{2},\mathbf{1})_3\oplus (\mathbf{2},\mathbf{1})_{-3}\oplus (\mathbf{1},\mathbf{3})_{-2}\oplus (\mathbf{1},\mathbf{\bar{3}})_{2}\oplus  (\mathbf{1},\mathbf{1})_0^{\oplus 2}$, then the theory fits into the $\mathbf{27}$ representation of $E_6$ (see e.g.~\cite{Gursey:1975ki,Hewett:1988xc}) and into the $\mathbf{15}\oplus \mathbf{\bar{6}}\oplus\mathbf{\bar{6}}$ representation of $\mathfrak{su}(6)\subset E_6$ (see e.g.~\cite{Fukugita:1981gn,Dutta:2016ach,Hartanto:2005jr}).

We can use this information to cross-check SuperFlocci. Doing so, we find $68$ semisimple extensions with $11$ of them being maximal. These are shown in Table~\ref{tab:E6Alg}. The output contains as expected $E_6$ as a maximal and $\mathfrak{su}(6)$ as a non-maximal extension.
\\~\\
\textbf{$\u$ extensions:} $\mathfrak{u}(1)$ extensions are important in BSM model building. If the $\mathfrak{u}(1)$ originates from a semisimple gauge algebra in the UV, non-trivial constraints on the model such as anomaly cancellation or the absence of Landau poles can be explained. One might therefore consider models with a semisimple UV completion better motivated from a theoretical point of view. Ref.~\cite{Davighi:2022dyq} found all of the $\mathfrak{u}(1)$ extensions of SM$_\nu$ that possess a semisimple UV completion. SuperFlocci complements these results by providing the ability to explicitly check the existence of a semisimple completion for $\mathfrak{u}(1)$ extension of any given theory. 

Reproducing some of the results of~\cite{Davighi:2022dyq} is an additional non-trivial check of SuperFlocci. As an example let us consider three generations of SM$_\nu$ with gauged baryon minus lepton number $B-L$~\cite{Davidson:1978pm,Mohapatra:1980qe}. The resulting $14$ maximal embeddings in Table~\ref{tab:BmL} agree with the ones in Table~4 of~\cite{Davighi:2022dyq}.\footnote{Note that Table~4 of~\cite{Davighi:2022dyq} gives only a subset of the $14$ algebras that we find. In order to complete the list one has to take into account equivalence classes of planes. Once this is done the outputs fully agree.} Analogously, the $\mathfrak{u}(1)$ extensions for which no semisimple completion was found in~\cite{Davighi:2022dyq} can be checked in SuperFlocci and we find perfect agreement.
%
%%%%%%%
\subsection{BSM models}
%%%%%%%
%
In addition to the above checks, we ran our code on a series of example BSM models. In each of these cases, novel semisimple extensions were found. 

A summary of the results is presented in Table~\ref{tab:BSMModels}. We now discuss some notable features. 

The addition of a singlet to the three-generation SM$_\nu$ is Model 1 of the list. 

 Despite having a $\u$ present, the LR-symmetric Models 2 and 3 both only have one minimal extension~ \cite{PhysRevD.11.2558,PhysRevD.12.1502}. This corresponds to Pati-Salam. The only other extension in the case of Model 2 is the $\so(10)$ GUT

Now compare Models 4, 5, 6 and 7, which are based on the $\su(5)$ GUT with a varying number of generations and number of RHNs. Note that adding a single RHN gives a bigger increase in the number of semisimple extensions than adding a new generation (comparing 5 and 6). 

Model 8 in the list corresponds to the fermionic representations of the minimal-supersymmetric standard model with RHNs (ignoring gauginos), and Model 9 to the 4-3-2-1 model~ \cite{Georgi:2016xhm,DiLuzio:2017vat}. Model 10, trinification~ \cite{Babu:1985gi,trini}, has two semisimple extensions corresponding to $E_6$ and $\su(6)$, the former being maximal and the latter minimal.

\begin{table*}
\renewcommand{\arraystretch}{1.5} 
\fontsize{8.5}{9.2}\selectfont
  \centering
  \begin{tabular}{|c|c| c|c|c|c|c|}
  \hline 
 & Model &  	Algebra & Representation & Total  &Max &  Min\\ \hline
  1& SM plus 1 RNH& 	$\su(2)\oplus\su(3)\oplus \u$ & $\text{SM}_\nu^{\oplus 3} \oplus (\mathbf{1},\mathbf{1},0)$ &587 & 34&5\\
   	  2  & 1-gen LR-symmetric&	$\su(2)\oplus \su(2) \oplus\su(3)\oplus \u$ &$(\mathbf{2},\mathbf{1},\mathbf{3},1)\oplus (\mathbf{1},\mathbf{2},\bar{\mathbf{3}},1)\oplus (\mathbf{2},\mathbf{1},\mathbf{1},-3)\oplus (\mathbf{1},\mathbf{2},\mathbf{1},3)$ & 2&1 &1\\
   	  	3& LR-symmetric&	  	$\su(2)\oplus \su(2) \oplus\su(3)\oplus \u$ & $[(\mathbf{2},\mathbf{1},\mathbf{3},1)\oplus (\mathbf{1},\mathbf{2},\bar{\mathbf{3}},1)\oplus (\mathbf{2},\mathbf{1},\mathbf{1},-3)\oplus (\mathbf{1},\mathbf{2},\mathbf{1},3)]^{\oplus 3}$ &173 &12 &1\\
   4&4-gen $\su(5)$&	  $\su(5)$ & $[\bar{\mathbf 5}\oplus \mathbf{10} ]^{\oplus 4}$ &156 &10 &1\\
  5&4-gen  $\su(5)$ with RHNs&		$\su(5)$ & $[\bar{\mathbf 5}\oplus \mathbf{10} \oplus \mathbf1]^{\oplus 4}$ &3146 & 38&1\\
  6&5-gen $\su(5)$&		$\su(5)$ & $[\bar{\mathbf 5}\oplus \mathbf{10} ]^{\oplus 5}$ &567 &20 &1\\
  7&5-gen  $\su(5)$ with RHNs&	$\su(5)$ & $[\bar{\mathbf 5}\oplus \mathbf{10} \oplus \mathbf 1]^{\oplus 5}$ & 22,820& 100&1\\
  8 & MSSM&	$\su(2)\oplus\su(3)\oplus \u$ & $\text{SM}_\nu^{\oplus 3} \oplus (\mathbf{2},\mathbf 1,-3)\oplus  (\mathbf{2},\mathbf1,3)$ &1337 & 40&7\\
  	9  &4-3-2-1&$\su(2)\oplus\su(3)\oplus \su(4) \oplus \u$ & $\text{SM}_\nu^{\oplus 2} \oplus (\mathbf{2},\mathbf{1},\mathbf{4},0)\oplus  (\mathbf{1},\mathbf{1},\bar{\mathbf{4}},-3)\oplus  (\mathbf{1},\mathbf{1},{\mathbf{4}},3) $ & 155& 11&3\\
  10& Trinification&	$\su(3)\oplus \su(3)\oplus \su(3)$& $(\mathbf{3},\bar{\mathbf{3}},\mathbf{1})\oplus(\bar{\mathbf{3}},\mathbf{3},\mathbf{1}) \oplus (\mathbf 1,\mathbf{3},\bar{\mathbf{3}})$ &2 & 1 & 1\\ \hline
  \end{tabular}
   \caption{Example BSM models run through SuperFlocci. The fourth column shows the total number of semisimple extensions, the fifth column shows the number of maximal algebras, and the last column shows the number of minimal algebras. Here SM$_\nu$ indicates the fermionic content of the one-generation SM with a RHN.}
  \label{tab:BSMModels}
\end{table*}
%
%%%%%%%
\subsection{Scans}
%%%%%%%
%
Arguably the most physically relevant application of SuperFlocci is to find semisimple extensions of the SM gauge algebra with a fermion spectrum that goes beyond that of the SM. For this reason, we performed a scan over $5835$ anomaly-free extensions of the SM$_\nu$ fermion spectrum and provide the results in~\cite{github}, such that the user can simply look up the semisimple extensions of their favorite model. In the scan, we assumed one generation of SM fermions including a RHN and added up to five additional BSM particles in representations $(\mathbf{a},\mathbf{b})_Y$ under $\mathfrak{su}(2)\oplus \mathfrak{su}(3)\oplus \mathfrak{u}(1)$, where $\mathbf{a}=\mathbf{1},\mathbf{2},\mathbf{3}$, $\mathbf{b} = \mathbf{1},\mathbf{3},\mathbf{6},\mathbf{8}$ with their conjugate representations and $-6 \leq Y \leq 6$. We additionally restrict the total dimension of the additional representations to be less than $30$.\footnote{Note that for $296$ models we were not able to perform the pruning as described in Section~\ref{sec:pruning} due to insufficient computational memory and processing time. We still provide the results after the steps outlined in Section~\ref{sec:grafting} for all of these models in a separate file and note that the resulting list of extensions is a superset of all true extensions, i.e. no extension is missed. If the user wishes to skip the pruning step due to memory or time issues, this can easily be achieved by setting the option \texttt{ExtendedKappaCheck} in the \texttt{SuperFlocci} function to False (see Table~\ref{tab:flocciargs}).}

We note that we did not perform the scan for three generations of SM$_\nu$ fermions since already for the SM$_\nu$ fermion content it takes approximately one hour to run, compared to a few seconds for one generation. However, we note that the one-generation results can also be viewed as the ``family universal" three-generation results.

The vast majority of maximal extensions have the same structure as the semisimple extensions for the SM$_\nu$ particle content, i.e. the SM gauge algebra is embedded within $\mathfrak{su}(5)$, $\mathfrak{su}(2)\oplus\mathfrak{su}(2)\oplus\mathfrak{su}(4)$, or $\mathfrak{so}(10)$. These often come with additional symmetries which account for the added fermions. However, there are a few exceptions. One of them is the $E_6$ model that we discussed in the previous section and which is shown in Table~\ref{tab:E6Alg}. Another is the occurrence of Pati-Salam type algebras of the form
\begin{equation}
	\mathfrak{su}(2)\oplus \mathfrak{su}(2)\oplus\mathfrak{su}(k)\,,\quad k\geq 4\,,
\end{equation}
with the fermions in the $(\mathbf{2},\mathbf{1},\mathbf{k})\oplus (\mathbf{1},\mathbf{2},\mathbf{\bar{k}})$ representation. These are only a few examples and the full result can be found in the GitHub repository.

%%%%%%%%%%%%%%%%%
\section{An example calculation: the one-generation SM}\label{sec:example}
%%%%%%%%%%%%%%%%%

In this section, we will explain the algorithm by working out an explicit example. We begin by specifying as an input 1) a finite-dimensional reductive Lie algebra $\r$ (the ``gauge algebra") and 2) the highest weights of a fully-reducible, $n$-dimensional representation $r:\r \to \su(n)$ (the ``fermionic representation"). The Lie algebra $\r$ can be decomposed into its semisimple and abelian parts as $\r = \s \oplus \a$. As our example, we will use the one-generation $\text{SM}_\nu$, that is, $\r = \su(2) \oplus \su(3) \oplus \u$ and the highest weights of $r$ are $Q\oplus U\oplus D\oplus L\oplus E\oplus N$, where these irreps\footnote{Strictly, these are equivalence classes of irreps, but we will simply refer to them as irreps.} were defined in Eq.~(\ref{eq:qudlen}). The dimension $n$ of this representation is $16$. Note that this was also the example used in Section~\ref{sec:running}.

\begin{figure}
    \centering
    \begin{minipage}{0.15\textwidth}
        \centering
        \begin{tikzcd}
        	& {\su(n)} \\
        	& {\g^\prime} &&& {} \\
        	& \g \\
        	\s & \r
        	\arrow["\kappa", from=4-2, to=3-2]
        	\arrow["\rho", from=3-2, to=2-2]
        	\arrow[from=4-1, to=4-2]
        	\arrow["\alpha", from=4-1, to=3-2]
        	\arrow["{\alpha^\prime}", curve={height=-12pt}, from=4-1, to=2-2]
        	\arrow["{\kappa^\prime}"', curve={height=12pt}, from=4-2, to=2-2]
        	\arrow["{\beta^\prime}"', from=2-2, to=1-2]
        \end{tikzcd}
    \end{minipage}%
    \begin{minipage}{0.24\textwidth}
        \centering
        \begin{tikzcd}
        	& {(\g^\prime,\beta^\prime)} \\
        	& {(\g,\beta)} \\
        	{(\s,r|_{\s})} & {(\r,r)}
        	\arrow["\Lambda\rho"', from=1-2, to=2-2]
        	\arrow["\Lambda\kappa"', from=2-2, to=3-2]
        	\arrow["{\Lambda \alpha}"', from=2-2, to=3-1]
        	\arrow[from=3-2, to=3-1]
        	\arrow["{\Lambda\kappa^\prime}", curve={height=-18pt}, from=1-2, to=3-2]
        	\arrow["{\Lambda\alpha^\prime}"', curve={height=18pt}, from=1-2, to=3-1]
        \end{tikzcd}
    \end{minipage}
    \caption{A diagram summarizing the projection matrices used within this paper (right), and their corresponding embeddings (left).}
    \label{fig:Maps}
\end{figure}
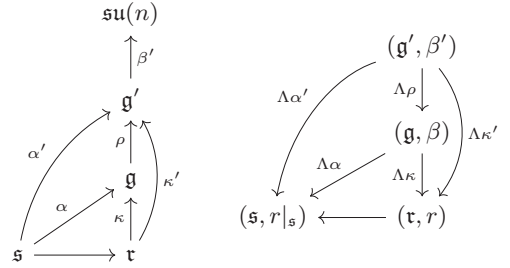

As discussed in Section~\ref{sec:output}, the object that we are trying to construct is a polytree, \emph{i.e} a directed acyclic graph consisting of vertices (or nodes) and edges (see Figure \ref{fig:aux} for an example). Every node consists of a semisimple Lie algebra $\g$, a representation $\beta$ of that algebra, and a projection matrix $\Lambda \kappa$ that encodes both the embedding of $\r$ into $\g$ (up to equivalence) and how $\beta$ branches down into $r$. This data is shown in Figure~\ref{fig:Maps}. Every edge (which we direct downwards) connects two nodes, which we may call the parent (upper) and child (lower) node, and consists of a projection matrix $\Lambda \rho$. This projection matrix loosely describes a maximal embedding of the child's algebra into the parent's algebra, and tells one how the parent's representation branches into the child's representation. That the embedding is maximal means that no other semisimple subalgebra of the parent's algebra contains the child's algebra. Finally, note that the graph is acyclic because the property of being a subalgebra is transitive.

At the base of the tree (the sinks) lie the minimal nodes (or ``minimal semisimple extensions"). If the input algebra is semisimple then the input node will be in the tree, and the sole minimal node.  Since all of the nodes in the tree must have algebras which are subalgebras of $\su (n)$, there must be maximal nodes. Those nodes that do not branch from any others in the tree are called maximal, and occupy the top nodes of the tree (sources of the tree).

The construction of the tree consists of three phases. First, an auxiliary tree is grown upward from the input algebra. The nodes of the auxiliary tree differ from those of the final tree in that $\Lambda \kappa$ is replaced with a projection matrix $\Lambda \alpha$ for only  $\s$, the semisimple part of $\r$ (see Figure~\ref{fig:Maps}). If the input's algebra is semisimple then we are done at this point. Second, the final tree is grafted onto the scaffolding of the auxiliary tree, and the projection matrices $\Lambda \kappa$ are constructed. Then, a final upward pruning is required to ensure that the $\Lambda \kappa$ translate into full algebra embeddings.

\subsection{Growing the auxiliary tree}

The algorithm generates the tree recursively, by creating all possible parent nodes for each existing node (we can think of $\s$ and $r|_{\s}$ as forming the base node of the auxiliary tree). These parent nodes then become child nodes in some future iteration. Firstly, for a given (child) node, a list of minimal superalgebras  (specified by an algebra and projection matrix $\Lambda \rho$) of the child's algebra is produced. Note that the same algebra may appear multiple times with different projection matrices.

In our example $\s = \su(2) \oplus \su(3)$. We will do the first step, and thus $\g=\s$. There are 10 minimal superalgebras of $\g$, with $8$ distinct algebras~$\g^\prime$:
\begin{center}
\begin{tabular}{ c p{0.2in} c }
 $\su(2)^{\oplus 2} \oplus \su(3)$ && $\su(2) \oplus \su(3)^{ \oplus 2}$\\ 
 $\su(3) \oplus \su(3)$ && $\su(3) \oplus \spalg(4)$\\  
 $\su(2) \oplus \su(4)$ && $\su(2) \oplus\su(6)$\\
 $\su(5)$ && $\su(6)$.
\end{tabular}
\end{center}
As an example of an algebra with more than one projection matrix to the base, take the algebra $\su(2)^{\oplus 2} \oplus \su(3)$. There is an embedding where the SM $\su(2)$ embeds via the identity into a single parent $\su(2)$ (say, the second), and an embedding where the SM $\su(2)$ embeds diagonally into both parent $\su(2)$ factors. Their projection matrices are given respectively by
\begin{equation}\label{eq:lambdarho}
    \Lambda \rho_1 = \begin{pmatrix}
    0 & 1 & 0 & 0\\
    0 & 0 & 1 & 0\\
    0 & 0 & 0 & 1
    \end{pmatrix} \quad \text{and} \quad
    \Lambda \rho_2 = \begin{pmatrix}
    1 & 1 & 0 & 0\\
    0 & 0 & 1 & 0\\
    0 & 0 & 0 & 1
    \end{pmatrix}
\end{equation}
Note how the matrix heights and widths reflect the ranks of the child's (in this case $\g=\s$ with rank $3$) and parent's (in this case $\su(2)^{\oplus 2} \oplus \su(3)$ with rank $4$) algebras respectively. This makes sense since the projection matrices map the parent weight system to the child weight system. The rows and columns of projection matrices are ordered according to our convention. For example, the first two columns of the matrices in Eq.~(\ref{eq:lambdarho}) correspond to the parent $\su(2)$ factors, while the last two correspond to the $\su(3)$ factor.

For each minimal superalgebra, there are 7 steps to find the representations that will form parent nodes. Each candidate may lead to zero, one, or several representations. We will follow the steps of the algorithm for the minimal superalgebra consisting of $\g^\prime=\su(2)^{\oplus 2} \oplus \su(3)$ and $\Lambda \rho_1$. The steps are as follows:\footnote{Strictly speaking, \textbf{U5} and \textbf{U6} are not required to form the auxiliary graph containing simply the embeddings of $\s$, since they involve the charges of $r$ under $\a$. However, if a node in the auxiliary tree fails to satisfy either of these criteria, the corresponding final tree nodes and their parents will not fulfil them either. Consequently, we greatly reduce computation time by adding these steps to the upward tree generation.}
\begin{itemize}
    \item[\bf U1:] Generate all irreps  (labelled by their unique highest weight) of the minimal superalgebra $\g'$ with dimension less than or equal to $n$ (in our case $16$). There are 85 such irreps.
    \item[\bf U2:] Keep only those highest weights whose projection under $\Lambda\rho$ is present in the weight system of the child representation.\footnote{Note that this step serves only to speed up the algorithm since the same irrep eliminations would be performed by {\bf U3}.} There are now 31 such irreps.
    \item[\bf U3:] Now keep only those irreps whose entire weight system, once projected down, is contained within the child weight system (taking weight multiplicities into account). There are now 6 such irreps: $(\mathbf{1},\mathbf{1},\mathbf{1})$, $(\mathbf{2},\mathbf{1},\mathbf{1})$, $(\mathbf{1},\mathbf{2},\mathbf{1})$, $(\mathbf{1},\mathbf{1},\mathbf{\overline{3}})$, $(\mathbf{2},\mathbf{1},\mathbf{\overline{3}})$, and $(\mathbf{1},\mathbf{2},\mathbf{3})$. Note that we included the singlet.
    \item[\bf U4:] From the remaining irreps, construct all representations of dimension exactly $n$ that project down to exactly the child weight system\footnote{Rather than being a set, the weight system is a multiset: the number of instances of each weight is important. The projection must not only produce the correct weights but also the correct multiplicity of each weight.} and satisfy anomaly cancellation (see e.g.~\cite{Yamatsu:2015npn}). For our case, there are 3 such representations:
\begin{center}
\begin{tabular}{ c }
 $(\mathbf{1},\mathbf{2},\mathbf{1}) \oplus (\mathbf{1},\mathbf{2},\mathbf{3}) \oplus (\mathbf{2},\mathbf{1},\mathbf{1}) \oplus (\mathbf{2},\mathbf{1},\mathbf{\overline{3}})$, \\ 
 $(\mathbf{1},\mathbf{2},\mathbf{1}) \oplus (\mathbf{1},\mathbf{2},\mathbf{3}) \oplus (\mathbf{2},\mathbf{1},\mathbf{1}) \oplus (\mathbf{1},\mathbf{1},\mathbf{\overline{3}})^{\oplus 2}$, \\  
 $(\mathbf{1},\mathbf{2},\mathbf{1}) \oplus (\mathbf{1},\mathbf{2},\mathbf{3}) \oplus (\mathbf{1},\mathbf{1},\mathbf{1})^{\oplus 2} \oplus (\mathbf{2},\mathbf{1},\mathbf{\overline{3}})$.
\end{tabular}
\end{center}
One immediately sees that each representation is composed of irreps that branch down to the left-handed leptons, left-handed quarks, right-handed leptons, and right-handed quarks, respectively.
    \item[\bf U5:] For each remaining $n$-dimensional representation, check whether it is possible to assign charges under $\a$ to each $\g^\prime$-weight such that 1) under projection the charges assigned to the $\s$-weights correspond exactly to the $\r$-weights of the input representation, and 2) the assignments are $\mathfrak{h}_{\g^\prime} \oplus \a$ anomaly free. Of the 3 representations in the last step, only the first passes this test.
	\item[\bf U6:] For each remaining $n$-dimensional representation, check that the weights of each irrep can be grouped into $\s$-irreps with constant $\a$ charge.
    \item[\bf U7:] Finally, the automorphisms of the minimal superalgebra are used to put the remaining representations which we now denote $\beta^\prime$ and projection matrices $\Lambda \rho$ and $\Lambda \alpha^\prime=\Lambda \alpha \circ \Lambda \rho$  into a unique form. This step is ultimately related to finding a representative of the equivalence classes representing nodes, as discussed in Section~\ref{sec:output}.
\end{itemize}

If a node, specified by $(\g^\prime, \beta^\prime, \Lambda \alpha^\prime)$ is not already present in the tree, it is added. However, the node may already be present. It is for this reason that the object we are constructing is not technically a tree, but rather a polytree, since branches can join back up with others.

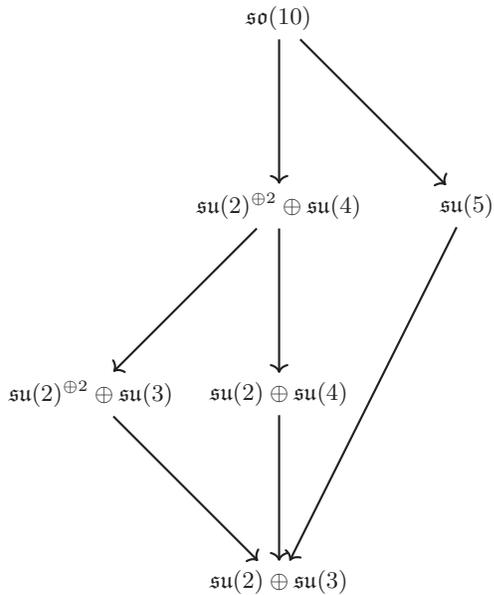
\begin{figure}[h]
\begin{tikzpicture}[node distance={25mm}, thick, main/.style = {}] 
\node[main] (1) {$\su(2) \oplus \su(3)$}; 
\node[main] (3) [above of=1] {$\su(2) \oplus \su(4)$}; 
\node[main] (2) [left of=3] {$\su(2)^{\oplus 2} \oplus \su(3)$}; 
\node[main] (5) [above of=3] {$\su(2)^{\oplus 2} \oplus \su(4)$}; 
\node[main] (4) [right of=5] {$\su(5)$}; 
\node[main] (6) [above of=5] {$\so(10)$}; 
\draw[->] (2) -- (1); 
\draw[->] (3) -- (1); 
\draw[->] (5) -- (2); 
\draw[->] (5) -- (3); 
\draw[->] (4) -- (1);
\draw[->] (6) -- (5); 
\draw[->] (6) -- (4);
\end{tikzpicture} 
\caption{The auxiliary tree of the one-generation SM$_\nu$.}
\label{fig:aux}
\end{figure}

Once no more nodes can be created, the auxiliary graph is complete. The graph for our example at this stage of the algorithm is given in Figure \ref{fig:aux}. If the input algebra is semisimple, this is the full graph and we stop.

\subsection{Grafting the final tree}\label{sec:grafting}

We must now produce a new graph from those nodes of the auxiliary graph, in which $\Lambda \alpha$ extends to embeddings of the full (reductive) input algebra into $\g$. Note that for a single semisimple embedding of the auxiliary tree, there may be multiple reductive embeddings. The algorithm begins at the top of the auxiliary tree and works downwards, with 4 steps being applied at each node.

\begin{itemize}
    \item[\bf D1:] The algorithm attempts to extends $\Lambda \alpha$ to a $\Lambda \kappa$ by adding extra rows for each $\u$ factor in the input algebra. The entries of these new rows are constrained so that $\beta$ projects to representations with the correct input $\u$ charges. There may be zero, one, or several solutions.
    \item[\bf D2:] It now checks which solutions are related by automorphisms of $\g$, and creates one node in the final tree for each equivalence class (see Section~\ref{sec:output}).
    \item[\bf D3:] If there are no solutions, it stops working on the current node of the auxiliary tree, and will skip all nodes below it in the auxiliary tree, which are guaranteed to have no solutions.
    \item[\bf D4:] For each newly created node (if any) in the final tree, check for all parents $p$ that $\Lambda \kappa_p = \Lambda \kappa \circ \Lambda \rho$, up to automorphism (see Section~\ref{sec:output}). If true, create an edge between them.
\end{itemize}

As an example, take the maximal algebra from the auxiliary tree, $\so(10)$. The steps of the last section yield
\begin{equation}
\Lambda \alpha = \begin{pmatrix}
    1 & 0 & 0 & 0 & 0\\
    0 & 0 & -1 & -1 & -1\\
    0 & 0 & 0 & 0 & 1
    \end{pmatrix}.
\end{equation}
One can check that the projection of the weight system given by the highest weight $(0,0,0,1,0)$ (the $\mathbf{16}$ irrep) yields the semisimple weights of the input representation. The full weights of the input representation (including $\u$ charges) can be obtained by extending $\Lambda \alpha$ with one of the two following rows:
\begin{center}
\begin{tabular}{ c }
 $(3,6,8,2,4)$, \\ 
 $(-3,-6,-4,-4,-2)$.
\end{tabular}
\end{center}
These two choices turn out to be related by automorphism, so one node in the final tree is created.

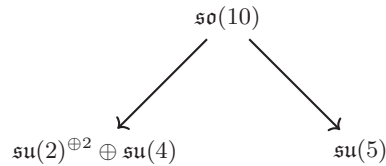
\begin{figure}[h]
\begin{tikzpicture}[node distance={25mm}, thick, main/.style = {}] 
\node[main] (1) {$\so(10)$}; 
\node[main] (2) [below left of=1] {$\su(2)^{\oplus 2} \oplus \su(4)$}; 
\node[main] (3) [below right of=1] {$\su(5)$}; 
\draw[->] (1) -- (2); 
\draw[->] (1) -- (3); 
\end{tikzpicture} 
\caption{The final tree of the one-generation SM$_\nu$. Notice the presence of the usual $\text{SM}_\nu$ extensions: $\so(10)$, Georgi-Glashow, and Pati-Salam. Their representations are $\mathbf{16}$, $\mathbf{10} \oplus \mathbf{\overline{5}} \oplus \mathbf{1}$, and $(\mathbf{2},\mathbf{1},\mathbf{4}) \oplus (\mathbf{1},\mathbf{2},\mathbf{\overline{4}})$, respectively.}
\label{fig:final}
\end{figure}

Once all nodes of the auxiliary tree have been processed, one has the final tree which ``sits on top" of the auxiliary tree. In general, some regions of the auxiliary tree will be absent in the final tree. At the same time, some auxiliary nodes will correspond to many final tree nodes. The graph at this stage is given in Figure \ref{fig:final}.

\subsection{Pruning the final tree}\label{sec:pruning}

The final step is to ensure the projection matrix implies the existence of an embedding (this is one of the conditions for a node in Section~\ref{sec:output}). This is true for semisimple algebras, but not in general for reductive algebras, though exceptions are not common. An exemplary input representation for the SM gauge algebra which produces projection matrices that do not correspond to an embedding is the one-generation $\text{SM}_\nu$ with additional vector-like fermions in the $(\mathbf{1},\mathbf{8})_4 \oplus (\mathbf{1},\mathbf{8})_{-4}$ representation.

The problem can arise when one attempts to extend the embedding of the Cartan subalgebra (given by the projection matrix) to an embedding of the full algebra. One parameterizes the embeddings of the non-Cartan generators with unknown parameters. Since commutation relations in the image must be satisfied, this gives a set of polynomial equations that is quadratic in these parameters. We do not actually need a solution, rather we simply need to know if a solution exists. We use a Gr\"obner basis analysis to carry this out.

All of the projection matrices in our example can be extended to full reductive embeddings. Therefore, the final tree takes the form of Figure \ref{fig:final}.

\section{Conclusion}

SuperFlocci is a versatile and user-friendly program to find all semisimple extensions to theories of arbitrary gauge algebra and (local) anomaly-free fermion representation. The program returns not only all semisimple extensions, but also detailed information about the embeddings of the input algebra into these extensions, the embeddings of extensions into each other, and the branching structure of their fermion representations. As an example application of the program, we performed a scan of 5835 extensions to the fermion content of the one-generation \SMnu. Beyond extending the fermion content of the SM, we believe SuperFlocci will be useful for those searching for GUTs of BSM theories with an extended gauge algebra.

\let\oldaddcontentsline\addcontentsline% Store \addcontentsline
\renewcommand{\addcontentsline}[3]{}% Make \addcontentsline a no-op
\section*{Acknowledgments}
	We thank Ben Allanach, Joe Davighi, Mijo Ghosh, Ben Gripaios, Alessandro Podo, and Eric San for helpful discussions.
 The authors are supported in part by the NSF grant PHY-2014071. AG is also supported by the NSERC PGS-D fellowship. MR is also supported by a Feodor-Lynen Research fellowship awarded by the Humboldt Foundation.
\let\addcontentsline\oldaddcontentsline% Restore \addcontentsline

\appendix

%%%%%%%%%%%%%%%%%%%
\section{Dynkin diagram convention}\label{app:DynkinDiagrams}
%%%%%%%%%%%%%%%%%%%
We follow the same convention in labelling our simple roots as in~\cite{Feger:2019tvk} (which is different from Flocci~\cite{Allanach:2021bfe}). As mentioned within the main text, all weights used within our program should be written within this convention.  
~\\
\tikzset{DynkinArrow/.style={shorten <=-1ex, shorten >=-1ex}}
\paragraph*{Dynkin diagram for $A_n$:}
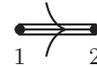
\begin{align*}
\begin{tikzpicture}[node distance={10mm}, thick, main/.style = {}] 
\node (1) [label=below:{\small $1$}] {};
\node (2) [label=below:{\small $2$},right of=1]  {};
\node (3) [label=below:{\small $n-1$},right of=2]  {};
\node (4) [label=below:{\small $n$},right of=3]  {};
\draw [DynkinArrow] (1)--(2);
\draw[dashed] (2)--(3);
\draw [DynkinArrow] (3)--(4);
\foreach \x in {(1), (2),(3),(4)}{
        \fill \x circle[radius=2pt];
    }
\end{tikzpicture} 
\end{align*}
%%%%%%%%%%%%%%%%%%%%%%%%%%
\paragraph*{Dynkin diagram for $B_n$:}
\begin{align*}
\begin{tikzpicture}[node distance={10mm}, thick, main/.style = {}] 
\node (1) [label=below:{\small $1$}] {};
\node (2) [label=below:{\small $2$},right of=1]  {};
\node (3) [label=below:{\small $n-1$},right of=2]  {};
\node (4) [label=below:{\small $n$},right of=3]  {};
\draw [DynkinArrow] (1)--(2);
\draw[dashed] (2)--(3);
  \draw[DynkinArrow,double distance=2pt, postaction={decorate,
                    decoration={markings,mark=at position 0. 75with {\arrow{>}}}
                   }] (3) -- (4);
\foreach \x in {(1), (2),(3),(4)}{
        \fill \x circle[radius=2pt];
    }
\end{tikzpicture} 
\end{align*}
%%%%%%%%%%%%%%%%%%%%%%%%%%
\paragraph*{Dynkin diagram for $C_n$:}
\begin{align*}
\begin{tikzpicture}[node distance={10mm}, thick, main/.style = {}] 
\node (1) [label=below:{\small $1$}] {};
\node (2) [label=below:{\small $2$},right of=1]  {};
\node (3) [label=below:{\small $n-1$},right of=2]  {};
\node (4) [label=below:{\small $n$},right of=3]  {};
\draw [DynkinArrow] (1)--(2);
\draw[DynkinArrow,dashed] (2)--(3);
  \draw[DynkinArrow,double distance=2pt,postaction={decorate,
                    decoration={markings,mark=at position 0. 75with {\arrow{<}}}
                   }] (3) -- (4);
\foreach \x in {(1), (2),(3),(4)}{
        \fill \x circle[radius=2pt];
    }
\end{tikzpicture} 
\end{align*}
%%%%%%%%%%%%%%%%%%%%%%%%%%
\paragraph*{Dynkin diagram for $D_n$:}
\begin{align*}
\begin{tikzpicture}[node distance={10mm}, thick, main/.style = {}] 
\node (1) [label=below:{\small $1$}] {};
\node (2) [label=below:{\small $2$},right of=1]  {};
\node (3) [label=right:{\small $n-2$},right of=2]  {};
\node (4) [label=below:{\small $n$},below right of=3]  {};
\node (5) [label={\small $n-1$},above right of=3]  {};
\draw [DynkinArrow] (1)--(2);
\draw [DynkinArrow] (3)--(4);
\draw [DynkinArrow] (3)--(5);
\draw[DynkinArrow,dashed] (2)--(3);
\foreach \x in {(1), (2),(3),(4),(5)}{
        \fill \x circle[radius=2pt];
    }
\end{tikzpicture} 
\end{align*}
%%%%%%%%%%%%%%%%%%%%%%%%%%
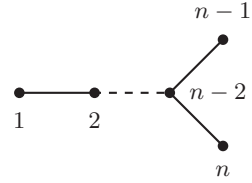
\paragraph*{Dynkin diagram for $E_6$:}
\begin{align*}
\begin{tikzpicture}[node distance={10mm}, thick, main/.style = {}] 
\node (1) [label=below:{\small $1$}] {};
\node (2) [label=below:{\small $2$},right of=1]  {};
\node (3) [label=below:{\small $3$},right of=2]  {};
\node (4) [label=below:{\small $4$},right of=3]  {};
\node (5) [label=below:{\small $5$},right of=4]  {};
\node (6) [label=right:{\small $6$},above of=3]  {};
\draw [DynkinArrow] (1)--(2);
\draw [DynkinArrow] (2)--(3);
\draw [DynkinArrow] (3)--(4);
\draw [DynkinArrow] (4)--(5);
\draw [DynkinArrow] (3)--(6);
\foreach \x in {(1), (2),(3),(4),(5),(6)}{
        \fill \x circle[radius=2pt];
    }
\end{tikzpicture} 
\end{align*}
%%%%%%%%%%%%%%%%%%%%%%%%%%
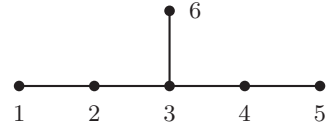
\paragraph*{Dynkin diagram for $E_7$:}
\begin{align*}
\begin{tikzpicture}[node distance={10mm}, thick, main/.style = {}] 
\node (1) [label=below:{\small $1$}] {};
\node (2) [label=below:{\small $2$},right of=1]  {};
\node (3) [label=below:{\small $3$},right of=2]  {};
\node (4) [label=below:{\small $4$},right of=3]  {};
\node (5) [label=below:{\small $5$},right of=4]  {};
\node (6) [label=below:{\small $6$},right of=5]  {};
\node (7) [label=right:{\small $7$},above of=3]  {};
\draw [DynkinArrow] (1)--(2);
\draw [DynkinArrow] (2)--(3);
\draw [DynkinArrow] (3)--(4);
\draw [DynkinArrow] (4)--(5);
\draw [DynkinArrow] (5)--(6);
\draw [DynkinArrow] (3)--(7);
\foreach \x in {(1), (2),(3),(4),(5),(6),(7)}{
        \fill \x circle[radius=2pt];
    }
\end{tikzpicture} 
\end{align*}
%%%%%%%%%%%%%%%%%%%%%%%%%%
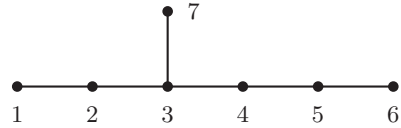
\paragraph*{Dynkin diagram for $E_8$:}
\begin{align*}
\begin{tikzpicture}[node distance={10mm}, thick, main/.style = {}] 
\node (1) [label=below:{\small $1$}] {};
\node (2) [label=below:{\small $2$},right of=1]  {};
\node (3) [label=below:{\small $3$},right of=2]  {};
\node (4) [label=below:{\small $4$},right of=3]  {};
\node (5) [label=below:{\small $5$},right of=4]  {};
\node (6) [label=below:{\small $6$},right of=5]  {};
\node (7) [label=below:{\small $7$},right of=6]  {};
\node (8) [label=right:{\small $8$},above of=3]  {};
\draw [DynkinArrow] (1)--(2);
\draw [DynkinArrow] (2)--(3);
\draw [DynkinArrow] (3)--(4);
\draw [DynkinArrow] (4)--(5);
\draw [DynkinArrow] (3)--(8);
\draw [DynkinArrow] (5)--(6);
\draw [DynkinArrow] (6)--(7);
\foreach \x in {(1), (2),(3),(4),(5),(6),(7),(8)}{
        \fill \x circle[radius=2pt];
    }
\end{tikzpicture} 
\end{align*}
%%%%%%%%%%%%%%%%%%%%%%%%%%
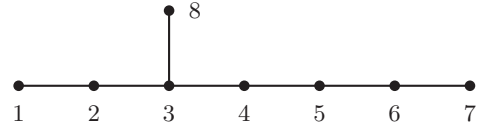
\paragraph*{Dynkin diagram for $F_4$:}
\begin{align*}
\begin{tikzpicture}[node distance={10mm}, thick, main/.style = {}] 
\node (1) [label=below:{\small $1$}] {};
\node (2) [label=below:{\small $2$},right of=1]  {};
\node (3) [label=below:{\small $3$},right of=2]  {};
\node (4) [label=below:{\small $4$},right of=3]  {};
  \draw[DynkinArrow,double distance=3pt, postaction={decorate,
                    decoration={markings,mark=at position 0. 75with {\arrow{>}}}
                   }] (2) -- (3);
\draw [DynkinArrow] (1)--(2);
\draw [DynkinArrow] (3)--(4);
\foreach \x in {(1), (2),(3),(4)}{
        \fill \x circle[radius=2pt];
    }
\end{tikzpicture} 
\end{align*}
%%%%%%%%%%%%%%%%%%%%%%%%%%
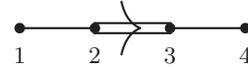
\paragraph*{Dynkin diagram for $G_2$:}
\begin{align*}
\begin{tikzpicture}[node distance={10mm}, thick, main/.style = {}] 
\node (1) [label=below:{\small $1$}] {};
\node (2) [label=below:{\small $2$},right of=1]  {};
  \draw[DynkinArrow,double distance=3pt, postaction={decorate,
                    decoration={markings,mark=at position 0. 75with {\arrow{>}}}
                   }] (1) -- (2);
\draw [DynkinArrow] (1)--(2);
\foreach \x in {(1), (2)}{
        \fill \x circle[radius=2pt];
    }
\end{tikzpicture} 
\end{align*}
%%%%%%%%%%%%%%%%%%%
\section{Tables}\label{app:tables}
%%%%%%%%%%%%%%%%%%%
%
This appendix contains tables generated by SuperFlocci for inputs specified in Section~\ref{sec:checkscan}.

\begin{table*}
\renewcommand{\arraystretch}{1.5} 
\fontsize{8}{9.2}\selectfont
\begin{tabular}{| c | r | l |} \hline
\multicolumn{3}{|c|}{\small Maximal } \\ \hline
  
 \hline
& {\small Algebra}&\small Fermion representations corresponding to $\beta$ \\ \hline 

%%%%%%%%%%%%%%%%%%%%%%%%%%%%%%%%%%%%%%%%%%%%%%%%%%%%%%%%
1 & $\mathfrak{su}(2)^{\oplus 2} \oplus \mathfrak{su}(5)$ & $(\mathbf{1} , \mathbf{3} , \mathbf{10})\oplus (\mathbf{3} , \mathbf{1} , \mathbf{\overline{5}})$ \\ \hline 
%%%%%%%%%%%%%%%%%%%%%%%%%%%%%%%%%%%%%%%%%%%%%%%%%%%%%%%%
2 & $\mathfrak{su}(2)^{\oplus 2} \oplus \mathfrak{su}(5)$ & $(\mathbf{1} , \mathbf{1} , \mathbf{5})\oplus (\mathbf{1} , \mathbf{2} , \mathbf{5})\oplus (\mathbf{3} , \mathbf{1} , \mathbf{\overline{10}})$ \\ \hline 
%%%%%%%%%%%%%%%%%%%%%%%%%%%%%%%%%%%%%%%%%%%%%%%%%%%%%%%%
3 & $\mathfrak{su}(2)^{\oplus 2} \oplus \mathfrak{su}(5)$ & $(\mathbf{1} , \mathbf{1} , \mathbf{10})\oplus (\mathbf{1} , \mathbf{2} , \mathbf{10})\oplus (\mathbf{3} , \mathbf{1} , \mathbf{\overline{5}})$ \\ \hline 
%%%%%%%%%%%%%%%%%%%%%%%%%%%%%%%%%%%%%%%%%%%%%%%%%%%%%%%%
4 & $\mathfrak{su}(5)^{\oplus 3}$ & $(\mathbf{1} , \mathbf{1} , \mathbf{\overline{5}})\oplus (\mathbf{1} , \mathbf{1} , \mathbf{10})\oplus (\mathbf{1} , \mathbf{\overline{5}} , \mathbf{1})\oplus (\mathbf{1} , \mathbf{10} , \mathbf{1})\oplus (\mathbf{\overline{5}} , \mathbf{1} , \mathbf{1})\oplus (\mathbf{10} , \mathbf{1} , \mathbf{1})$ \\ \hline 
%%%%%%%%%%%%%%%%%%%%%%%%%%%%%%%%%%%%%%%%%%%%%%%%%%%%%%%%
5 & $\mathfrak{su}(2)^{\oplus 2} \oplus \mathfrak{su}(5)^{\oplus 2}$ & $(\mathbf{1} , \mathbf{1} , \mathbf{1} , \mathbf{\overline{5}})\oplus (\mathbf{1} , \mathbf{1} , \mathbf{1} , \mathbf{10})\oplus (\mathbf{1} , \mathbf{2} , \mathbf{10} , \mathbf{1})\oplus (\mathbf{2} , \mathbf{1} , \mathbf{\overline{5}} , \mathbf{1})$ \\ \hline 
%%%%%%%%%%%%%%%%%%%%%%%%%%%%%%%%%%%%%%%%%%%%%%%%%%%%%%%%
%%%%%%%%%%%%%%%%%%%%%%%%%%%%%%%%%%%%%%%%%%%%%%%%%%%%%%%%
\multicolumn{3}{|c|}{\small Minimal } \\ \hline%%%%%%%%%%%%%%%%%%%%%%%%%%%%%%%%%%%%%%%%%%%%%%%%%%%%%%%%
%%%%%%%%%%%%%%%%%%%%%%%%%%%%%%%%%%%%%%%%%%%%%%%%%%%%%%%%
%%%%%%%%%%%%%%%%%%%%%%%%%%%%%%%%%%%%%%%%%%%%%%%%%%%%%%%%
6 & $\mathfrak{su}(5)$ & $\mathbf{\overline{5}}^{\oplus 3}\oplus \mathbf{10}^{\oplus 3}$ \\ \hline 
%%%%%%%%%%%%%%%%%%%%%%%%%%%%%%%%%%%%%%%%%%%%%%%%%%%%%%%%
\end{tabular}
\caption{\label{tab:MinMaxSMwoRHN} All maximal and minimal algebras for three generations of SM fermions without RHNs. The gauge algebra is $\mathfrak{su}(2) \oplus \mathfrak{su}(3) \oplus \mathfrak{u}(1)$ and the input representation $[(\mathbf{2} , \mathbf{3} , 1)\oplus (\mathbf{1} , \mathbf{\overline{3}} , -4)\oplus (\mathbf{1} , \mathbf{\overline{3}} , 2)\oplus (\mathbf{2} , \mathbf{1} , -3)\oplus (\mathbf{1} , \mathbf{1} , 6)]^{\oplus 3}$.}
\end{table*}
%%%%%%%%%%%%%%%%%%%%%%%%%%%%%%%%%%%%%%%%%%%%%%%%%%%%%%%%

\begin{table*}
\renewcommand{\arraystretch}{1.5} 
\fontsize{8}{9.2}\selectfont
\begin{tabular}{| c | r | l |} \hline
\multicolumn{3}{|c|}{\small Maximal } \\ \hline
& {\small Algebra}&\small Fermion representations corresponding to $\beta$ \\ \hline 

%%%%%%%%%%%%%%%%%%%%%%%%%%%%%%%%%%%%%%%%%%%%%%%%%%%%%%%%
1 & $\mathfrak{su}(2) \oplus \mathfrak{so}(10)$ & $(\mathbf{3} , \mathbf{16})$ \\ \hline 
%%%%%%%%%%%%%%%%%%%%%%%%%%%%%%%%%%%%%%%%%%%%%%%%%%%%%%%%
2 & $\mathfrak{su}(2) \oplus \mathfrak{so}(10)^{\oplus 2}$ & $(\mathbf{1} , \mathbf{1} , \mathbf{16})\oplus (\mathbf{2} , \mathbf{16} , \mathbf{1})$ \\ \hline 
%%%%%%%%%%%%%%%%%%%%%%%%%%%%%%%%%%%%%%%%%%%%%%%%%%%%%%%%
3 & $\mathfrak{so}(10)^{\oplus 3}$ & $(\mathbf{1} , \mathbf{1} , \mathbf{16})\oplus (\mathbf{1} , \mathbf{16} , \mathbf{1})\oplus (\mathbf{16} , \mathbf{1} , \mathbf{1})$ \\ \hline 
%%%%%%%%%%%%%%%%%%%%%%%%%%%%%%%%%%%%%%%%%%%%%%%%%%%%%%%%
4 & $\mathfrak{su}(4) \oplus \mathfrak{sp}(6)^{\oplus 2}$ & $(\mathbf{\overline{4}} , \mathbf{6} , \mathbf{1})\oplus (\mathbf{4} , \mathbf{1} , \mathbf{6})$ \\ \hline 
%%%%%%%%%%%%%%%%%%%%%%%%%%%%%%%%%%%%%%%%%%%%%%%%%%%%%%%%
5 & $\mathfrak{su}(2)^{\oplus 2} \oplus \mathfrak{su}(12)$ & $(\mathbf{1} , \mathbf{2} , \mathbf{12})\oplus (\mathbf{2} , \mathbf{1} , \mathbf{\overline{12}})$ \\ \hline 
%%%%%%%%%%%%%%%%%%%%%%%%%%%%%%%%%%%%%%%%%%%%%%%%%%%%%%%%
6 & $\mathfrak{su}(2)^{\oplus 3} \oplus \mathfrak{su}(4) \oplus \mathfrak{sp}(6)$ & $(\mathbf{1} , \mathbf{1} , \mathbf{1} , \mathbf{4} , \mathbf{6})\oplus (\mathbf{1} , \mathbf{1} , \mathbf{2} , \mathbf{\overline{4}} , \mathbf{1})\oplus (\mathbf{2} , \mathbf{2} , \mathbf{1} , \mathbf{\overline{4}} , \mathbf{1})$ \\ \hline 
%%%%%%%%%%%%%%%%%%%%%%%%%%%%%%%%%%%%%%%%%%%%%%%%%%%%%%%%
7 & $\mathfrak{su}(2)^{\oplus 3} \oplus \mathfrak{su}(4)^{\oplus 2}$ & $(\mathbf{1} , \mathbf{1} , \mathbf{1} , \mathbf{4} , \mathbf{6})\oplus (\mathbf{1} , \mathbf{1} , \mathbf{2} , \mathbf{\overline{4}} , \mathbf{1})\oplus (\mathbf{2} , \mathbf{2} , \mathbf{1} , \mathbf{\overline{4}} , \mathbf{1})$ \\ \hline 
%%%%%%%%%%%%%%%%%%%%%%%%%%%%%%%%%%%%%%%%%%%%%%%%%%%%%%%%
8 & $\mathfrak{su}(4)^{\oplus 2} \oplus \mathfrak{sp}(6)$ & $(\mathbf{\overline{4}} , \mathbf{6} , \mathbf{1})\oplus (\mathbf{4} , \mathbf{1} , \mathbf{6})$ \\ \hline 
%%%%%%%%%%%%%%%%%%%%%%%%%%%%%%%%%%%%%%%%%%%%%%%%%%%%%%%%
9 & $\mathfrak{su}(2)^{\oplus 3} \oplus \mathfrak{su}(4) \oplus \mathfrak{sp}(6)$ & $(\mathbf{1} , \mathbf{1} , \mathbf{1} , \mathbf{4} , \mathbf{6})\oplus (\mathbf{1} , \mathbf{1} , \mathbf{2} , \mathbf{\overline{4}} , \mathbf{1})\oplus (\mathbf{2} , \mathbf{2} , \mathbf{1} , \mathbf{\overline{4}} , \mathbf{1})$ \\ \hline 
%%%%%%%%%%%%%%%%%%%%%%%%%%%%%%%%%%%%%%%%%%%%%%%%%%%%%%%%
10 & $\mathfrak{su}(2)^{\oplus 2} \oplus \mathfrak{su}(4) \oplus \mathfrak{sp}(4) \oplus \mathfrak{so}(10)$ & $(\mathbf{1} , \mathbf{1} , \mathbf{1} , \mathbf{1} , \mathbf{16})\oplus (\mathbf{1} , \mathbf{1} , \mathbf{4} , \mathbf{4} , \mathbf{1})\oplus (\mathbf{2} , \mathbf{2} , \mathbf{\overline{4}} , \mathbf{1} , \mathbf{1})$ \\ \hline 
%%%%%%%%%%%%%%%%%%%%%%%%%%%%%%%%%%%%%%%%%%%%%%%%%%%%%%%%
11 & $\mathfrak{su}(2)^{\oplus 4} \oplus \mathfrak{su}(4) \oplus \mathfrak{so}(10)$ & $(\mathbf{1} , \mathbf{1} , \mathbf{1} , \mathbf{1} , \mathbf{1} , \mathbf{16})\oplus (\mathbf{1} , \mathbf{1} , \mathbf{2} , \mathbf{2} , \mathbf{4} , \mathbf{1})\oplus (\mathbf{2} , \mathbf{2} , \mathbf{1} , \mathbf{1} , \mathbf{\overline{4}} , \mathbf{1})$ \\ \hline 
%%%%%%%%%%%%%%%%%%%%%%%%%%%%%%%%%%%%%%%%%%%%%%%%%%%%%%%%
12 & $\mathfrak{su}(2)^{\oplus 2} \oplus \mathfrak{su}(8) \oplus \mathfrak{so}(10)$ & $(\mathbf{1} , \mathbf{1} , \mathbf{1} , \mathbf{16})\oplus (\mathbf{1} , \mathbf{2} , \mathbf{8} , \mathbf{1})\oplus (\mathbf{2} , \mathbf{1} , \mathbf{\overline{8}} , \mathbf{1})$ \\ \hline 
%%%%%%%%%%%%%%%%%%%%%%%%%%%%%%%%%%%%%%%%%%%%%%%%%%%%%%%%
13 & $\mathfrak{su}(4) \oplus \mathfrak{sp}(4)^{\oplus 2} \oplus \mathfrak{so}(10)$ & $(\mathbf{1} , \mathbf{1} , \mathbf{1} , \mathbf{16})\oplus (\mathbf{\overline{4}} , \mathbf{4} , \mathbf{1} , \mathbf{1})\oplus (\mathbf{4} , \mathbf{1} , \mathbf{4} , \mathbf{1})$ \\ \hline 
%%%%%%%%%%%%%%%%%%%%%%%%%%%%%%%%%%%%%%%%%%%%%%%%%%%%%%%%
14 & $\mathfrak{su}(2)^{\oplus 2} \oplus \mathfrak{su}(4) \oplus \mathfrak{sp}(4) \oplus \mathfrak{so}(10)$ & $(\mathbf{1} , \mathbf{1} , \mathbf{1} , \mathbf{1} , \mathbf{16})\oplus (\mathbf{1} , \mathbf{1} , \mathbf{4} , \mathbf{4} , \mathbf{1})\oplus (\mathbf{2} , \mathbf{2} , \mathbf{\overline{4}} , \mathbf{1} , \mathbf{1})$ \\ \hline 
%%%%%%%%%%%%%%%%%%%%%%%%%%%%%%%%%%%%%%%%%%%%%%%%%%%%%%%%
%%%%%%%%%%%%%%%%%%%%%%%%%%%%%%%%%%%%%%%%%%%%%%%%%%%%%%%%
\multicolumn{3}{|c|}{\small Minimal } \\ \hline%%%%%%%%%%%%%%%%%%%%%%%%%%%%%%%%%%%%%%%%%%%%%%%%%%%%%%%%
%%%%%%%%%%%%%%%%%%%%%%%%%%%%%%%%%%%%%%%%%%%%%%%%%%%%%%%%
%%%%%%%%%%%%%%%%%%%%%%%%%%%%%%%%%%%%%%%%%%%%%%%%%%%%%%%%
15 & $\mathfrak{su}(2) \oplus \mathfrak{su}(4)^{\oplus 2}$ & $(\mathbf{1} , \mathbf{4} , \mathbf{6})\oplus (\mathbf{2} , \mathbf{\overline{4}} , \mathbf{1})^{\oplus 3}$ \\ \hline 
%%%%%%%%%%%%%%%%%%%%%%%%%%%%%%%%%%%%%%%%%%%%%%%%%%%%%%%%
16 & $\mathfrak{su}(2)^{\oplus 2} \oplus \mathfrak{su}(4)$ & $(\mathbf{1} , \mathbf{2} , \mathbf{4})^{\oplus 3}\oplus (\mathbf{2} , \mathbf{1} , \mathbf{\overline{4}})^{\oplus 3}$ \\ \hline 
%%%%%%%%%%%%%%%%%%%%%%%%%%%%%%%%%%%%%%%%%%%%%%%%%%%%%%%%
\end{tabular}
\caption{All maximal and minimal algebras for three generations of SM$_\nu$ extended by a gauged $\mathfrak{u}(1)_{B-L}$. The gauge algebra is $\mathfrak{su}(2) \oplus \mathfrak{su}(3) \oplus \mathfrak{u}(1)\oplus \u$ and the input representation $[(\mathbf{2} , \mathbf{3} , 1,1)\oplus (\mathbf{1} , \mathbf{\overline{3}} , -4,-1)\oplus (\mathbf{1} , \mathbf{\overline{3}} , 2,-1)\oplus (\mathbf{2} , \mathbf{1} , -3,-3)\oplus (\mathbf{1} , \mathbf{1} , 6,3)\oplus(\mathbf{1} , \mathbf{1} , 0,3) ]^{\oplus 3}$.}
\label{tab:BmL}
\end{table*}
%%%%%%%%%%%%%%%%%%%%%%%%%%%%%%%%%%%%%%%%%%%%%%%%%%%%%%%%

\begin{table*}
\renewcommand{\arraystretch}{1.5} 
\fontsize{8}{9.2}\selectfont
\begin{tabular}{| c | r | l |} \hline
\multicolumn{3}{|c|}{\small Maximal } \\ \hline
& {\small Algebra}&\small Fermion representations corresponding to $\beta$ \\ \hline 

%%%%%%%%%%%%%%%%%%%%%%%%%%%%%%%%%%%%%%%%%%%%%%%%%%%%%%%%
1 & $\mathfrak{su}(2)^{\oplus 2} \oplus \mathfrak{su}(5)$ & $(\mathbf{1} , \mathbf{1} , \mathbf{10})\oplus (\mathbf{1} , \mathbf{1} , \mathbf{5})\oplus (\mathbf{1} , \mathbf{2} , \mathbf{1})\oplus (\mathbf{2} , \mathbf{1} , \mathbf{\overline{5}})$ \\ \hline 
%%%%%%%%%%%%%%%%%%%%%%%%%%%%%%%%%%%%%%%%%%%%%%%%%%%%%%%%
2 & $E_6$ & $\mathbf{27}$ \\ \hline 
%%%%%%%%%%%%%%%%%%%%%%%%%%%%%%%%%%%%%%%%%%%%%%%%%%%%%%%%
3 & $\mathfrak{su}(2) \oplus \mathfrak{su}(5) \oplus \mathfrak{so}(10)$ & $(\mathbf{1} , \mathbf{1} , \mathbf{10})\oplus (\mathbf{1} , \mathbf{\overline{5}} , \mathbf{1})\oplus (\mathbf{1} , \mathbf{10} , \mathbf{1})\oplus (\mathbf{2} , \mathbf{1} , \mathbf{1})$ \\ \hline 
%%%%%%%%%%%%%%%%%%%%%%%%%%%%%%%%%%%%%%%%%%%%%%%%%%%%%%%%
4 & $\mathfrak{sp}(10) \oplus \mathfrak{so}(10)$ & $(\mathbf{1} , \mathbf{1})\oplus (\mathbf{1} , \mathbf{16})\oplus (\mathbf{10} , \mathbf{1})$ \\ \hline 
%%%%%%%%%%%%%%%%%%%%%%%%%%%%%%%%%%%%%%%%%%%%%%%%%%%%%%%%
5 & $\mathfrak{su}(5) \oplus \mathfrak{sp}(12)$ & $(\mathbf{1} , \mathbf{12})\oplus (\mathbf{\overline{5}} , \mathbf{1})\oplus (\mathbf{10} , \mathbf{1})$ \\ \hline 
%%%%%%%%%%%%%%%%%%%%%%%%%%%%%%%%%%%%%%%%%%%%%%%%%%%%%%%%
6 & $\mathfrak{su}(5) \oplus \mathfrak{so}(12)$ & $(\mathbf{1} , \mathbf{12})\oplus (\mathbf{\overline{5}} , \mathbf{1})\oplus (\mathbf{10} , \mathbf{1})$ \\ \hline 
%%%%%%%%%%%%%%%%%%%%%%%%%%%%%%%%%%%%%%%%%%%%%%%%%%%%%%%%
7 & $\mathfrak{so}(11) \oplus \mathfrak{so}(10)$ & $(\mathbf{1} , \mathbf{16})\oplus (\mathbf{11} , \mathbf{1})$ \\ \hline 
%%%%%%%%%%%%%%%%%%%%%%%%%%%%%%%%%%%%%%%%%%%%%%%%%%%%%%%%
8 & $\mathfrak{su}(2)^{\oplus 2} \oplus \mathfrak{su}(5) \oplus \mathfrak{sp}(8)$ & $(\mathbf{1} , \mathbf{1} , \mathbf{1} , \mathbf{8})\oplus (\mathbf{1} , \mathbf{1} , \mathbf{\overline{5}} , \mathbf{1})\oplus (\mathbf{1} , \mathbf{1} , \mathbf{10} , \mathbf{1})\oplus (\mathbf{2} , \mathbf{2} , \mathbf{1} , \mathbf{1})$ \\ \hline 
%%%%%%%%%%%%%%%%%%%%%%%%%%%%%%%%%%%%%%%%%%%%%%%%%%%%%%%%
9 & $\mathfrak{su}(4) \oplus \mathfrak{su}(5) \oplus \mathfrak{sp}(6)$ & $(\mathbf{1} , \mathbf{1} , \mathbf{6})\oplus (\mathbf{1} , \mathbf{\overline{5}} , \mathbf{1})\oplus (\mathbf{1} , \mathbf{10} , \mathbf{1})\oplus (\mathbf{6} , \mathbf{1} , \mathbf{1})$ \\ \hline 
%%%%%%%%%%%%%%%%%%%%%%%%%%%%%%%%%%%%%%%%%%%%%%%%%%%%%%%%
10 & $\mathfrak{su}(4) \oplus \mathfrak{su}(5) \oplus \mathfrak{sp}(6)$ & $(\mathbf{1} , \mathbf{1} , \mathbf{6})\oplus (\mathbf{1} , \mathbf{\overline{5}} , \mathbf{1})\oplus (\mathbf{1} , \mathbf{10} , \mathbf{1})\oplus (\mathbf{6} , \mathbf{1} , \mathbf{1})$ \\ \hline 
%%%%%%%%%%%%%%%%%%%%%%%%%%%%%%%%%%%%%%%%%%%%%%%%%%%%%%%%
11 & $\mathfrak{sp}(4) \oplus \mathfrak{sp}(6) \oplus \mathfrak{so}(10)$ & $(\mathbf{1} , \mathbf{1} , \mathbf{16})\oplus (\mathbf{1} , \mathbf{6} , \mathbf{1})\oplus (\mathbf{5} , \mathbf{1} , \mathbf{1})$ \\ \hline 
%%%%%%%%%%%%%%%%%%%%%%%%%%%%%%%%%%%%%%%%%%%%%%%%%%%%%%%%
%%%%%%%%%%%%%%%%%%%%%%%%%%%%%%%%%%%%%%%%%%%%%%%%%%%%%%%%
\multicolumn{3}{|c|}{\small Minimal } \\ \hline%%%%%%%%%%%%%%%%%%%%%%%%%%%%%%%%%%%%%%%%%%%%%%%%%%%%%%%%
%%%%%%%%%%%%%%%%%%%%%%%%%%%%%%%%%%%%%%%%%%%%%%%%%%%%%%%%
%%%%%%%%%%%%%%%%%%%%%%%%%%%%%%%%%%%%%%%%%%%%%%%%%%%%%%%%
12 & $\mathfrak{su}(3)^{\oplus 3}$ & $(\mathbf{1} , \mathbf{3} , \mathbf{3})\oplus (\mathbf{\overline{3}} , \mathbf{\overline{3}} , \mathbf{1})\oplus (\mathbf{3} , \mathbf{1} , \mathbf{\overline{3}})$ \\ \hline 
%%%%%%%%%%%%%%%%%%%%%%%%%%%%%%%%%%%%%%%%%%%%%%%%%%%%%%%%
13 & $\mathfrak{su}(5) \oplus \mathfrak{sp}(6)^{\oplus 2}$ & $(\mathbf{1} , \mathbf{1} , \mathbf{6})\oplus (\mathbf{1} , \mathbf{6} , \mathbf{1})\oplus (\mathbf{\overline{5}} , \mathbf{1} , \mathbf{1})\oplus (\mathbf{10} , \mathbf{1} , \mathbf{1})$ \\ \hline 
%%%%%%%%%%%%%%%%%%%%%%%%%%%%%%%%%%%%%%%%%%%%%%%%%%%%%%%%
14 & $\mathfrak{su}(2)^{\oplus 2} \oplus \mathfrak{su}(5) \oplus \mathfrak{sp}(6)$ & $(\mathbf{1} , \mathbf{1} , \mathbf{1} , \mathbf{1})^{\oplus 2}\oplus (\mathbf{1} , \mathbf{1} , \mathbf{1} , \mathbf{6})\oplus (\mathbf{1} , \mathbf{1} , \mathbf{\overline{5}} , \mathbf{1})\oplus (\mathbf{1} , \mathbf{1} , \mathbf{10} , \mathbf{1})\oplus (\mathbf{2} , \mathbf{2} , \mathbf{1} , \mathbf{1})$ \\ \hline 
%%%%%%%%%%%%%%%%%%%%%%%%%%%%%%%%%%%%%%%%%%%%%%%%%%%%%%%%
15 & $\mathfrak{su}(2)^{\oplus 2} \oplus \mathfrak{su}(4) \oplus \mathfrak{su}(5)$ & $(\mathbf{1} , \mathbf{1} , \mathbf{1} , \mathbf{1})\oplus (\mathbf{1} , \mathbf{1} , \mathbf{1} , \mathbf{\overline{5}})\oplus (\mathbf{1} , \mathbf{1} , \mathbf{1} , \mathbf{5})\oplus (\mathbf{1} , \mathbf{2} , \mathbf{4} , \mathbf{1})\oplus (\mathbf{2} , \mathbf{1} , \mathbf{\overline{4}} , \mathbf{1})$ \\ \hline 
%%%%%%%%%%%%%%%%%%%%%%%%%%%%%%%%%%%%%%%%%%%%%%%%%%%%%%%%
16 & $\mathfrak{su}(2)^{\oplus 2} \oplus \mathfrak{su}(4) \oplus \mathfrak{su}(5)$ & $(\mathbf{1} , \mathbf{1} , \mathbf{1} , \mathbf{1})^{\oplus 2}\oplus (\mathbf{1} , \mathbf{1} , \mathbf{1} , \mathbf{\overline{5}})\oplus (\mathbf{1} , \mathbf{1} , \mathbf{1} , \mathbf{10})\oplus (\mathbf{1} , \mathbf{1} , \mathbf{6} , \mathbf{1})\oplus (\mathbf{2} , \mathbf{2} , \mathbf{1} , \mathbf{1})$ \\ \hline 
%%%%%%%%%%%%%%%%%%%%%%%%%%%%%%%%%%%%%%%%%%%%%%%%%%%%%%%%
17 & $\mathfrak{su}(5)$ & $\mathbf{1}^{\oplus 2}\oplus \mathbf{\overline{5}}^{\oplus 2}\oplus \mathbf{10}\oplus \mathbf{5}$ \\ \hline 
%%%%%%%%%%%%%%%%%%%%%%%%%%%%%%%%%%%%%%%%%%%%%%%%%%%%%%%%
18 & $\mathfrak{su}(2)^{\oplus 2} \oplus \mathfrak{su}(4) \oplus \mathfrak{sp}(6)$ & $(\mathbf{1} , \mathbf{1} , \mathbf{1} , \mathbf{1})\oplus (\mathbf{1} , \mathbf{1} , \mathbf{1} , \mathbf{6})\oplus (\mathbf{1} , \mathbf{2} , \mathbf{4} , \mathbf{1})\oplus (\mathbf{2} , \mathbf{1} , \mathbf{\overline{4}} , \mathbf{1})\oplus (\mathbf{2} , \mathbf{2} , \mathbf{1} , \mathbf{1})$ \\ \hline 
%%%%%%%%%%%%%%%%%%%%%%%%%%%%%%%%%%%%%%%%%%%%%%%%%%%%%%%%
19 & $\mathfrak{su}(2)^{\oplus 2} \oplus \mathfrak{su}(4)$ & $(\mathbf{1} , \mathbf{1} , \mathbf{1})\oplus (\mathbf{1} , \mathbf{1} , \mathbf{6})\oplus (\mathbf{1} , \mathbf{2} , \mathbf{4})\oplus (\mathbf{2} , \mathbf{1} , \mathbf{\overline{4}})\oplus (\mathbf{2} , \mathbf{2} , \mathbf{1})$ \\ \hline 
%%%%%%%%%%%%%%%%%%%%%%%%%%%%%%%%%%%%%%%%%%%%%%%%%%%%%%%%
\end{tabular}
\caption{All maximal and minimal algebras for one generation of SM fermions plus fermions in the vector-like $(\mathbf{2},\mathbf{1})_3\oplus (\mathbf{2},\mathbf{1})_{-3}\oplus (\mathbf{1},\mathbf{3})_{-2}\oplus (\mathbf{1},\mathbf{\bar{3}})_{2}\oplus  (\mathbf{1},\mathbf{1})_0^{\oplus 2}$ representation of $\mathfrak{su}(2)\oplus\mathfrak{su}(3)\oplus\mathfrak{u}(1)$.}
\label{tab:E6Alg}
\end{table*}

\let\oldaddcontentsline\addcontentsline% Store \addcontentsline
\renewcommand{\addcontentsline}[3]{}% Make \addcontentsline a no-op
%%%%%%%%%%%%%%%%%%%%%%
\bibliography{SuperFlocci}

\providecommand{\href}[2]{#2}\begingroup\raggedright\begin{thebibliography}{10}

\bibitem{Georgi:1974sy}
H.~Georgi and S.~L. Glashow, \emph{{Unity of All Elementary Particle Forces}},
  \href{http://dx.doi.org/10.1103/PhysRevLett.32.438}{\emph{Phys. Rev. Lett.}
  {\bf 32} (1974) 438--441}.

\bibitem{Fritzsch:1974nn}
H.~Fritzsch and P.~Minkowski, \emph{{Unified Interactions of Leptons and
  Hadrons}}, \href{http://dx.doi.org/10.1016/0003-4916(75)90211-0}{\emph{Annals
  Phys.} {\bf 93} (1975) 193--266}.

\bibitem{Georgi:1974my}
H.~Georgi, \emph{{The State of the Art\textemdash{}Gauge Theories}},
  \href{http://dx.doi.org/10.1063/1.2947450}{\emph{AIP Conf. Proc.} {\bf 23}
  (1975) 575--582}.

\bibitem{Pati:1974yy}
J.~C. Pati and A.~Salam, \emph{{Lepton Number as the Fourth Color}},
  \href{http://dx.doi.org/10.1103/PhysRevD.10.275}{\emph{Phys. Rev. D} {\bf 10}
  (1974) 275--289}.

\bibitem{Davighi:2022fer}
J.~Davighi and J.~Tooby-Smith, \emph{{Electroweak flavour unification}},
  \href{http://dx.doi.org/10.1007/JHEP09(2022)193}{\emph{JHEP} {\bf 09} (2022)
  193}, [\href{http://arxiv.org/abs/2201.07245}{{\tt 2201.07245}}].

\bibitem{Allanach:2021bfe}
B.~C. Allanach, B.~Gripaios and J.~Tooby-Smith, \emph{{Semisimple extensions of
  the Standard Model gauge algebra}},
  \href{http://dx.doi.org/10.1103/PhysRevD.104.035035}{\emph{Phys. Rev. D} {\bf
  104} (2021) 035035}, [\href{http://arxiv.org/abs/2104.14555}{{\tt
  2104.14555}}].

\bibitem{de2000lie}
W.~de~Graaf, \emph{Lie Algebras: Theory and Algorithms}.
\newblock ISSN. Elsevier Science, 2000.

\bibitem{Lorente:1972xw}
M.~Lorente and B.~Gruber, \emph{{Classification of semisimple subalgebras of
  simple lie algebras}}, \href{http://dx.doi.org/10.1063/1.1665888}{\emph{J.
  Math. Phys.} {\bf 13} (1972) 1639--1663}.

\bibitem{gruber1975semisimple}
B.~Gruber and M.~Samuel, \emph{Semisimple subalgebras of semisimple lie
  algebras: the algebra (su (6)) as a physically significant example},  in
  \emph{Group theory and its applications}, pp.~95--141.
\newblock Elsevier, 1975.

\bibitem{dyn52:MaximalSubgroups}
E.~B. Dynkin, \emph{Maximal subgroups of the classical groups}, {\emph{Tr.
  Mosk. Mat. Obs.} {\bf 1} (1952) 39--166}.

\bibitem{dynkin1952semisimple}
E.~Dynkin, \emph{{Semisimple subalgebras of semisimple Lie algebras}},
  {\emph{Matematicheskii Sbornik} {\bf 72} (1952) 349--462}.

\bibitem{Feger:2019tvk}
R.~Feger, T.~W. Kephart and R.~J. Saskowski, \emph{{LieART 2.0 \textendash{} A
  Mathematica application for Lie Algebras and Representation Theory}},
  \href{http://dx.doi.org/10.1016/j.cpc.2020.107490}{\emph{Comput. Phys.
  Commun.} {\bf 257} (2020) 107490},
  [\href{http://arxiv.org/abs/1912.10969}{{\tt 1912.10969}}].

\bibitem{Tooby-Smith:2021hjk}
J.~Tooby-Smith, \emph{{Arithmetical, geometrical, and categorical forays into
  particle physics}}.
\newblock PhD thesis, Cambridge U., 5, 2021.
\newblock 10.17863/CAM.72061.

\bibitem{github}
A.~Gomes, M.~Ruhdorfer and J.~Tooby-Smith, \emph{Superflocci github
  repository},  2023.

\bibitem{Gursey:1975ki}
F.~Gursey, P.~Ramond and P.~Sikivie, \emph{{A Universal Gauge Theory Model
  Based on E6}},
  \href{http://dx.doi.org/10.1016/0370-2693(76)90417-2}{\emph{Phys. Lett. B}
  {\bf 60} (1976) 177--180}.

\bibitem{Hewett:1988xc}
J.~L. Hewett and T.~G. Rizzo, \emph{{Low-Energy Phenomenology of Superstring
  Inspired E(6) Models}},
  \href{http://dx.doi.org/10.1016/0370-1573(89)90071-9}{\emph{Phys. Rept.} {\bf
  183} (1989) 193}.

\bibitem{Fukugita:1981gn}
M.~Fukugita, T.~Yanagida and M.~Yoshimura, \emph{{N anti-N OSCILLATION WITHOUT
  LEFT-RIGHT SYMMETRY}},
  \href{http://dx.doi.org/10.1016/0370-2693(82)91092-9}{\emph{Phys. Lett. B}
  {\bf 109} (1982) 369--372}.

\bibitem{Dutta:2016ach}
B.~Dutta, Y.~Gao, T.~Ghosh, I.~Gogoladze, T.~Li and J.~W. Walker, \emph{{SU(6)
  GUT origin of the TeV-scale vectorlike particles associated with the 750 GeV
  diphoton resonance}},
  \href{http://dx.doi.org/10.1103/PhysRevD.94.036006}{\emph{Phys. Rev. D} {\bf
  94} (2016) 036006}, [\href{http://arxiv.org/abs/1604.07838}{{\tt
  1604.07838}}].

\bibitem{Hartanto:2005jr}
A.~Hartanto and L.~T. Handoko, \emph{{Grand unified theory based on the SU(6)
  symmetry}}, \href{http://dx.doi.org/10.1103/PhysRevD.71.095013}{\emph{Phys.
  Rev. D} {\bf 71} (2005) 095013},
  [\href{http://arxiv.org/abs/hep-ph/0504280}{{\tt hep-ph/0504280}}].

\bibitem{Davighi:2022dyq}
J.~Davighi and J.~Tooby-Smith, \emph{{Flatland: abelian extensions of the
  Standard Model with semi-simple completions}},
  \href{http://dx.doi.org/10.1007/JHEP09(2022)159}{\emph{JHEP} {\bf 09} (2022)
  159}, [\href{http://arxiv.org/abs/2206.11271}{{\tt 2206.11271}}].

\bibitem{Davidson:1978pm}
A.~Davidson, \emph{{$B-L$ as the fourth color within an $\mathrm{SU}(2)_L
  \times \mathrm{U}(1)_R \times \mathrm{U}(1)$ model}},
  \href{http://dx.doi.org/10.1103/PhysRevD.20.776}{\emph{Phys. Rev. D} {\bf 20}
  (1979) 776}.

\bibitem{Mohapatra:1980qe}
R.~N. Mohapatra and R.~E. Marshak, \emph{{Local B-L Symmetry of Electroweak
  Interactions, Majorana Neutrinos and Neutron Oscillations}},
  \href{http://dx.doi.org/10.1103/PhysRevLett.44.1316}{\emph{Phys. Rev. Lett.}
  {\bf 44} (1980) 1316--1319}.

\bibitem{PhysRevD.11.2558}
R.~N. Mohapatra and J.~C. Pati, \emph{"natural" left-right symmetry},
  \href{http://dx.doi.org/10.1103/PhysRevD.11.2558}{\emph{Phys. Rev. D} {\bf
  11} (May, 1975) 2558--2561}.

\bibitem{PhysRevD.12.1502}
G.~Senjanovic and R.~N. Mohapatra, \emph{Exact left-right symmetry and
  spontaneous violation of parity},
  \href{http://dx.doi.org/10.1103/PhysRevD.12.1502}{\emph{Phys. Rev. D} {\bf
  12} (Sep, 1975) 1502--1505}.

\bibitem{Georgi:2016xhm}
H.~Georgi and Y.~Nakai, \emph{{Diphoton resonance from a new strong force}},
  \href{http://dx.doi.org/10.1103/PhysRevD.94.075005}{\emph{Phys. Rev. D} {\bf
  94} (2016) 075005}, [\href{http://arxiv.org/abs/1606.05865}{{\tt
  1606.05865}}].

\bibitem{DiLuzio:2017vat}
L.~Di~Luzio, A.~Greljo and M.~Nardecchia, \emph{{Gauge leptoquark as the origin
  of B-physics anomalies}},
  \href{http://dx.doi.org/10.1103/PhysRevD.96.115011}{\emph{Phys. Rev. D} {\bf
  96} (2017) 115011}, [\href{http://arxiv.org/abs/1708.08450}{{\tt
  1708.08450}}].

\bibitem{Babu:1985gi}
K.~S. Babu, X.-G. He and S.~Pakvasa, \emph{{Neutrino Masses and Proton Decay
  Modes in SU(3) X SU(3) X SU(3) Trinification}},
  \href{http://dx.doi.org/10.1103/PhysRevD.33.763}{\emph{Phys. Rev. D} {\bf 33}
  (1986) 763}.

\bibitem{trini}
A.~de~Rújula, H.~Georgi and S.~L. Glashow, \emph{Fifth workshop on grand
  unification},  (Singapore), p.~88, World Scientific, 1984.

\bibitem{Yamatsu:2015npn}
N.~Yamatsu, \emph{{Finite-Dimensional Lie Algebras and Their Representations
  for Unified Model Building}},  \href{http://arxiv.org/abs/1511.08771}{{\tt
  1511.08771}}.

\end{thebibliography}\endgroup
\let\addcontentsline\oldaddcontentsline% Restore \addcontentsline
\end{document}